\documentclass[reprint,amsmath,amssymb,aps,superscriptaddress]{revtex4-2}

\usepackage{graphicx}
\usepackage{dcolumn}
\usepackage{amsmath}
\usepackage{bm}
\usepackage{mathrsfs}
\usepackage{txfonts}
\usepackage{epstopdf}
\usepackage{subfigure}
\usepackage{caption}
\usepackage{float}

\begin{document}
	
\title{Disorder-induced phase transitions in double Weyl semimetals}

\author{Jiayan Zhang}
\affiliation{Department of Physics, Hangzhou Dianzi University, Hangzhou, Zhejiang 310018, China}

\author{Fei Wan}
\affiliation{Department of Physics, Hangzhou Dianzi University, Hangzhou, Zhejiang 310018, China}

\author{Xinru Wang}
\affiliation{Department of Physics, Hangzhou Dianzi University, Hangzhou, Zhejiang 310018, China}

\author{Ying Ding}
\affiliation{Department of Physics, Hangzhou Dianzi University, Hangzhou, Zhejiang 310018, China}

\author{Liehong Liao}
\affiliation{Department of Physics, Hangzhou Dianzi University, Hangzhou, Zhejiang 310018, China}

\author{Zhihui Chen}
\affiliation{Department of Physics, Hangzhou Dianzi University, Hangzhou, Zhejiang 310018, China}

\author{M. N. Chen}
\email[]{mnchen@hdu.edu.cn}

\author{Yuan Li}
\email[]{liyuan@hdu.edu.cn}
\affiliation{Department of Physics, Hangzhou Dianzi University, Hangzhou, Zhejiang 310018, China}
\date{\today}

\begin{abstract}
The double Weyl semimetal (DWSM) is a newly proposed topological material that hosts Weyl points with chiral charge $n$=2. The disorder effect in DWSM is investigated by adopting the tight-binding Hamiltonian. Using the transfer matrix method and the noncommutative Kubo formula, we numerically calculate the localization length and the Hall conductivity in the presence of the on-site nonmagnetic disorder or orbital (spin-flip) disorders, and give the corresponding global phase diagrams. For the on-site nonmagnetic disorder, the system undergoes the DWSM-3D quantum anomalous hall (3D QAH) and normal insulator (NI)-DWSM phase transitions, and evolves into the diffusive metal (DM) phase before being localized by strong disorders, which is consistent with the Weyl semimetal. For $\sigma_x$ orbital disorder, we find that increasing disorder can generate a pair of Weyl nodes at the boundary of the Brillouin zone and induce a 3D QAH-DWSM phase transition. Then we investigate the combined effect of orbital disorders for both disordered 3D QAH phase and DWSM phase. The disorder-induced transitions can be well understood in terms of an effective medium theory based on self-consistent Born approximation.
\end{abstract}


\maketitle

\section{Introduction}
Weyl semimetal is a three-dimensional topological state of matter, in which the conduction and valence bands touch at a finite number of nodes~\cite{Hasan,Armitage,Hasan2}. The Weyl nodes always appear in pairs and each Weyl node can be regarded as a monopole in $k$-space carrying the chiral charge $n$=1. Weyl semimetal has the Fermi arc surface states that connect the surface projections of two Weyl nodes~\cite{Wan}. Weyl semimetal has been predicted to exist in many materials~\cite{Wan,Burkov,Bulmash,Xu,Liu,Huang,Weng}. And researchers found the Weyl fermions in TaAs~\cite{SYX}, ferromagnetic semimetal $\rm Co_3Sn_2S_2$~\cite{Morali} and $\rm MoTe_2$~\cite{Deng}. Weyl semimetal supports many fascinating properties such as the chiral anomaly~\cite{Nielsen,HXC,Zyuzin,ZCL,ZJH}, quantum Hall effect~\cite{WCM,LHL,TFD}, anomalous Hall effect~\cite{EL,TL}, nonlinear optical effect~\cite{Morimoto,WL} and magneto-optical response~\cite{Polatkan,Okamura}.

However, the chiral charge of the Weyl node can be more than one, namely $n$$>$1, and the corresponding materials are named as multi-Weyl semimetals~\cite{FC,Huang3,CWJ,Ahn}. For $n$=2, which is double Weyl semimetal (DWSM), the dispersion relation in the vicinity of such node is quadratic in two directions and linear in the third direction. These Weyl nodes are protected by the crystallographic point group symmetries~\cite{FC}. The DWSM is theoretically proposed in $\rm HgCr_2Se_4$~\cite{Xu} and $\rm SrSi_2$~\cite{Huang3} and can be achieved in photonic crystals~\cite{CWJ}. Numerical calculations suggest the presence of multiple surface Fermi arcs in multi-Weyl semimetal~\cite{Umer,MXY}. The topological semimetal with a large chiral charge can also induce intriguing transport phenomena~\cite{Ahn}.

Quantum interference can completely suppress the diffusion of a particle in a random potential, a phenomenon known as Anderson localization~\cite{Anderson}. Surprisingly, Li et al. investigated the role of disorder in HgTe/CdTe quantum well and found that disorder can induce the topological phase transition and generate the topological Anderson insulator phase~\cite{LJ}. Afterward, the disorder effect is extensively discussed in 3D topological insulator~\cite{GHM}, Kane-Mele model~\cite{Orth}, HgTe/CdTe quantum well~\cite{JH,CCZ}, Dirac semimetal thin film~\cite{CR}, quasicrystal~\cite{CR2,PT} and Weyl semimetal~\cite{CCZ2,Shapourian,SY2,WYJ,Trescher,CR3}. The disorder can also induce phase transition between type-I and type-II Weyl semimetal~\cite{Park2}.  However, the role of disorder in DWSM hasn't been thoroughly investigated and needs to be further explored. Besides, most of the works only involve the on-site nonmagnetic disorder referring to the $\sigma_0$ term in the Hamiltonian. There also exists orbital disorder referring to the $\sigma_x$, $\sigma_y$ or $\sigma_z$ terms in the Hamiltonian. The orbital disorder can also alter the topological properties of various systems~\cite{CR3,SJT,HHH,QZH,HLH,YH}. In addition, the effect of the orbital disorder in DWSM has not been discussed yet.

In this paper, we study the combined effect of the on-site nonmagnetic and orbital disorders in DWSM by calculating the localization length and the Hall conductivity. These disorders can give rise to rich phase transitions shown in the phase diagrams. Firstly, the tight-binding Hamiltonian of DWSM is introduced in momentum and real space. And the phase diagram in the clean limit is presented. In the presence of disorder, the results indicate that the 3D quantum anomalous Hall (3D QAH) phase and DWSM phase are stable in the weak disorder. And the system undergoes a series of phase transitions with increasing of the disorder strength. For nonmagnetic disorder, phase transitions in DWSM are consistent with those reported in Weyl semimetal~\cite{CCZ2,Shapourian}. However, for the $\sigma_x$ orbital disorder, it is the first time for us to find a new phase transition from 3D QAH to DWSM phase. Besides, there is a directly phase transition from DWSM to normal insulator (NI), which is quite different from the effect of the nonmagnetic disorder. For the latter, the system must enter the diffusive metal (DM) phase before being localized by the strong disorder.  The combined effect of orbital disorders are also investigated for disordered 3D QAH and DWSM phases. Furthermore, an effective medium theory based on self-consistent Born approximation is introduced to explain the disorder-induced phase transitions. At last, we discuss the experimental realization of disordered DWSM.

The paper is organized as follows: In Sec.~\ref{sec:Method}, we introduce the model Hamiltonian and give the phase diagram in the clean limit. The numerical calculation methods are also given here. In Sec.~\ref{sec:Results}, we show the localization length and Hall conductivity and plot the global phase diagrams in the presence of disorder. In Sec.~\ref{sec:SCBA}, we interpret the disorder effect in terms of the self-consistent Born approximation (SCBA). Finally, a brief discussion and summary are given in Sec.~\ref{sec:Discussion}.

\section{MODEL AND METHOD}\label{sec:Method}

\begin{figure}[htbp]
\includegraphics[scale=0.26]{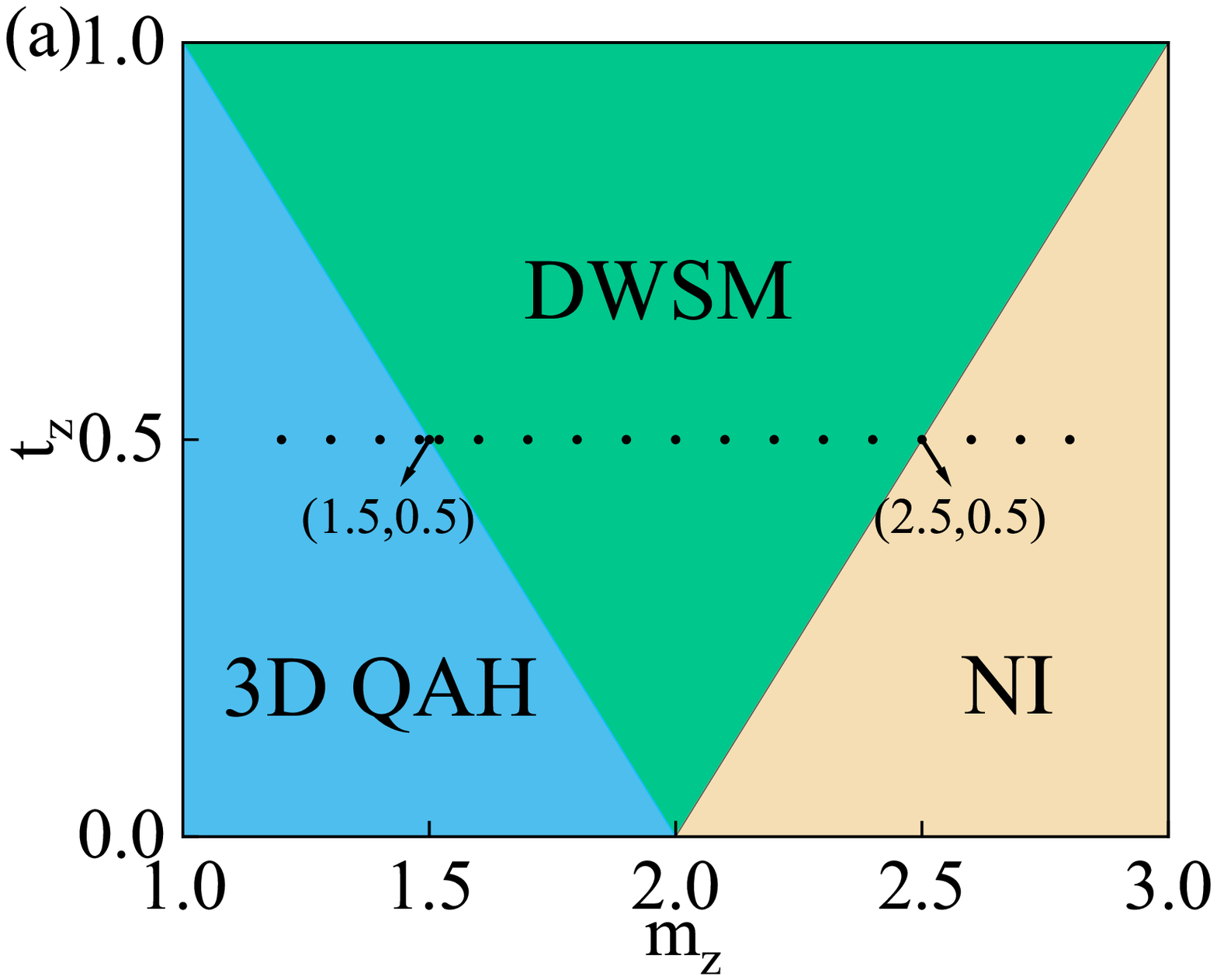}
\includegraphics[scale=0.22]{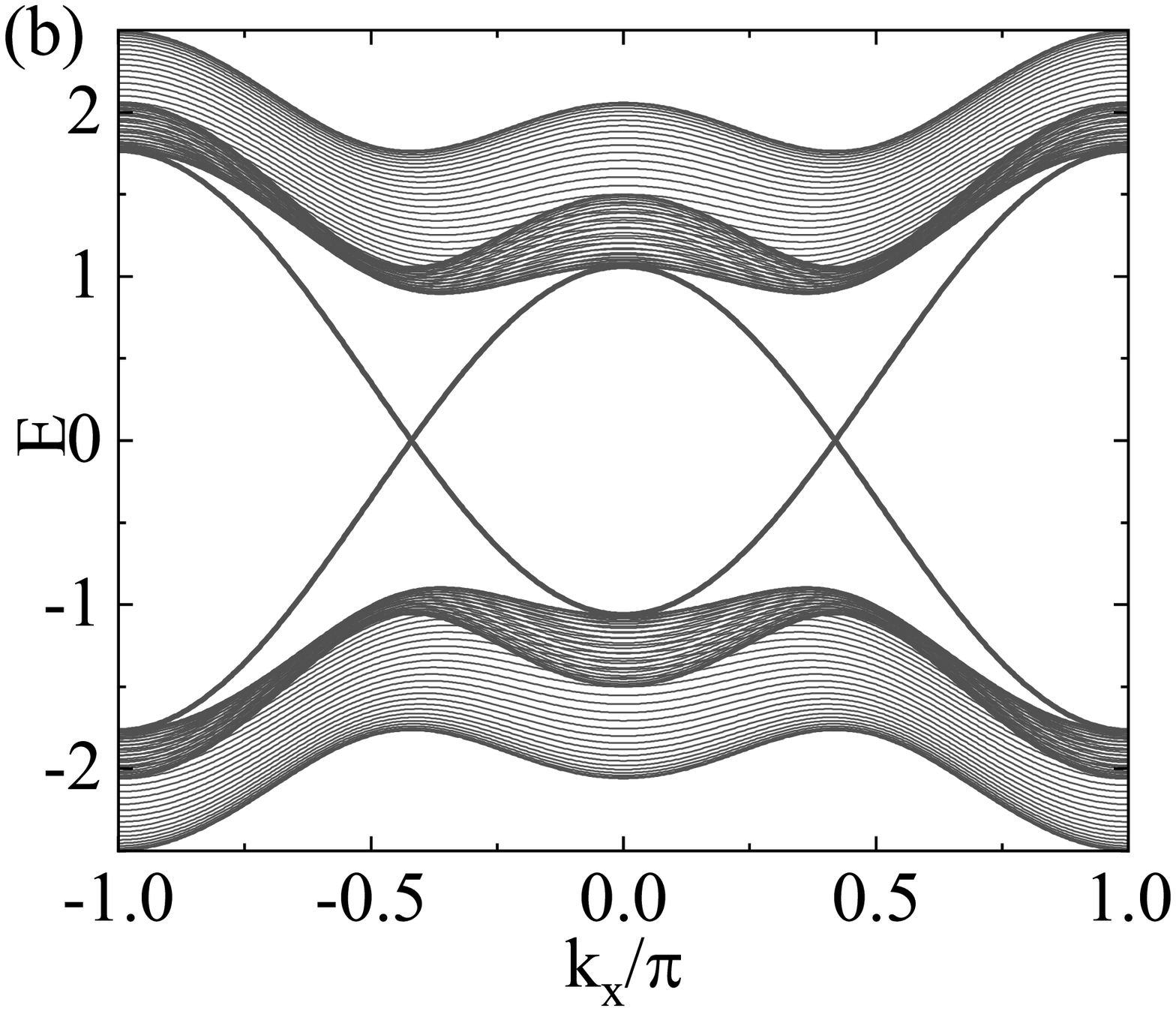}

\includegraphics[scale=0.22]{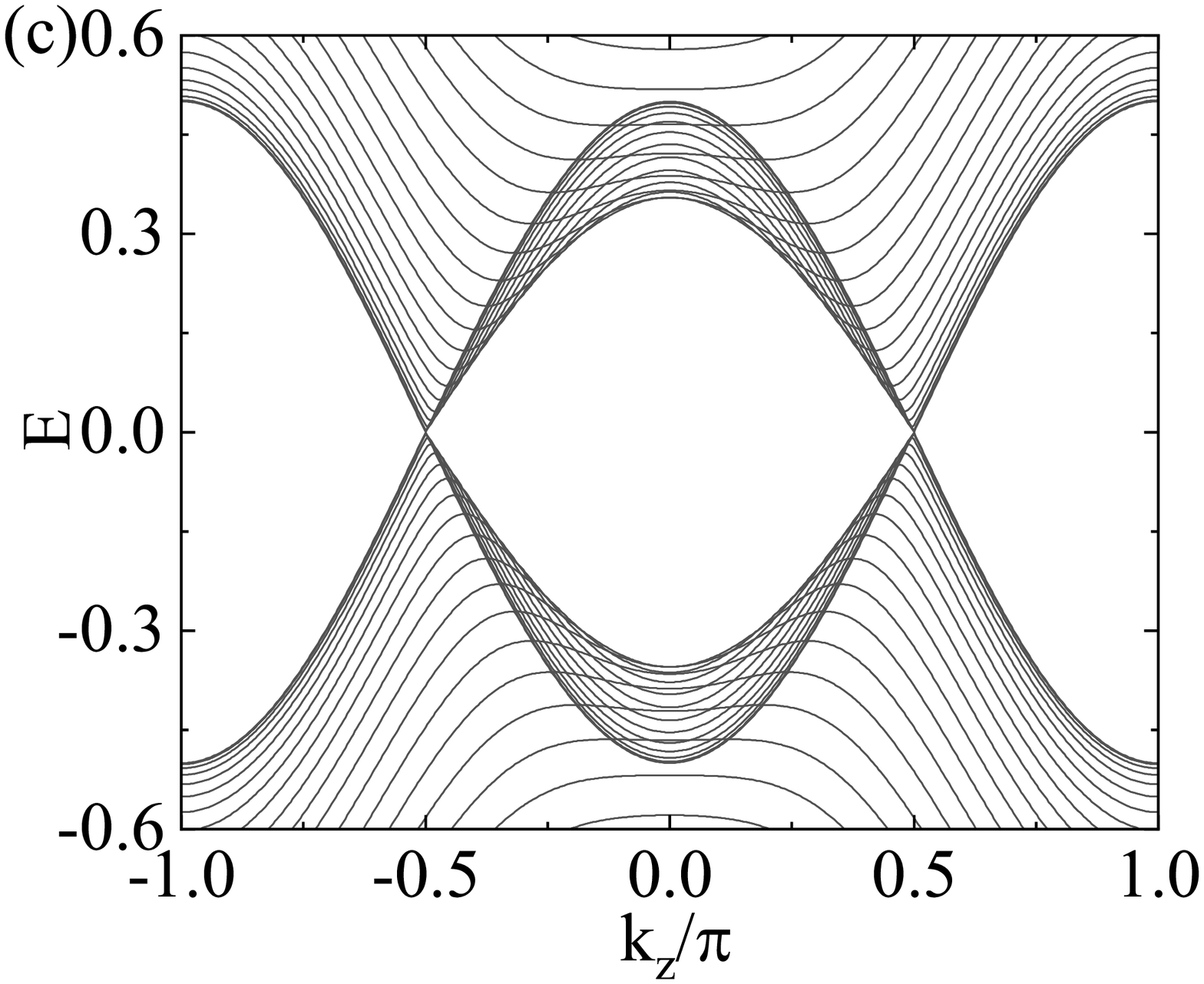}
\includegraphics[scale=0.22]{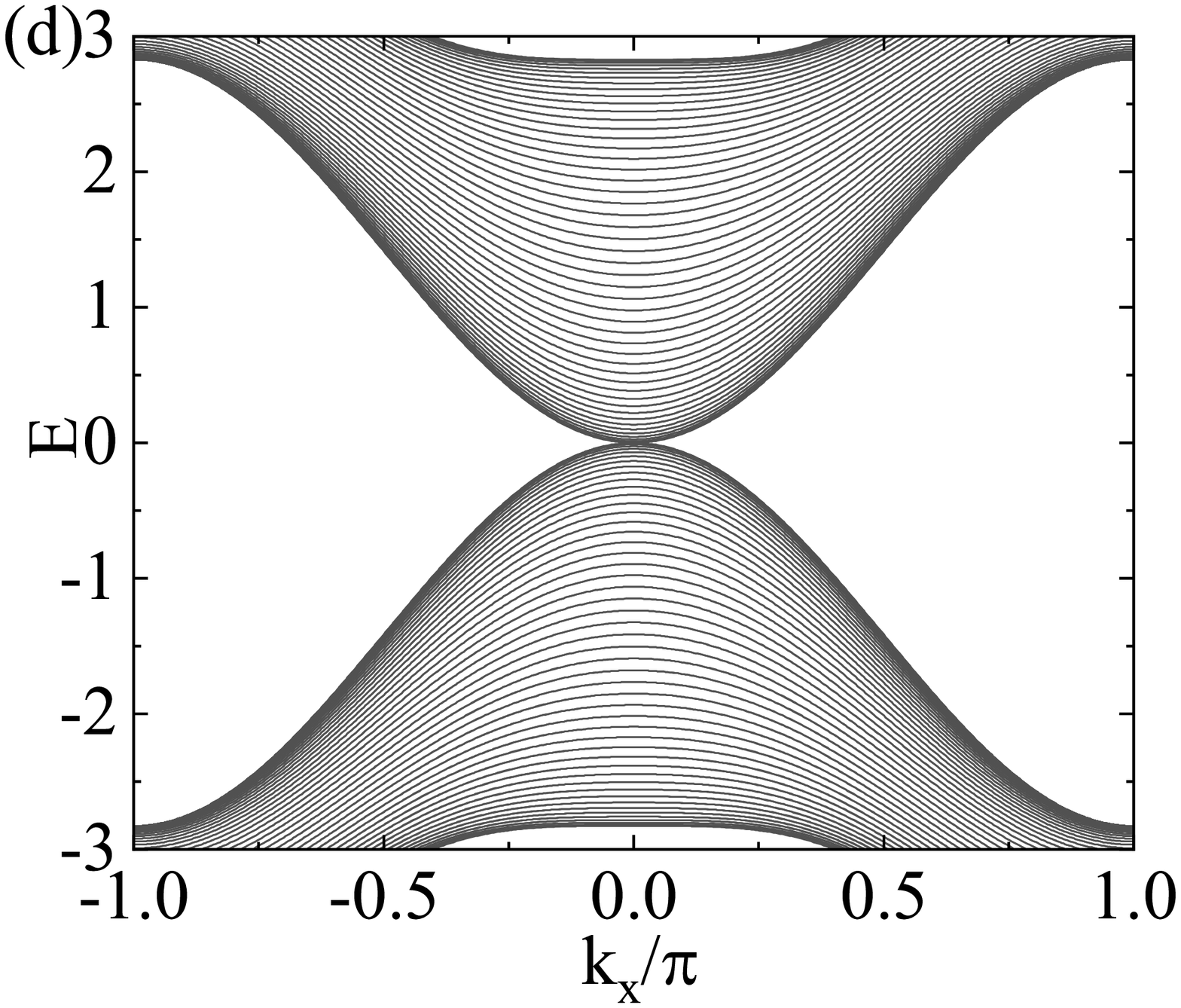}
\caption{\label{figure1}(a) Phase diagram of DWSM without disorder on the $t_z$-$m_z$ plane. There are 3D QAH, DWSM and NI phase for different $m_z$. (b) Band structure of 3D QAH phase with open boundary condition in the $y$ direction ($N_y$=50) for $k_z$=0 and $m_z$=1. (c) Band structure on the $E$-$k_z$ plane with $k_x$=$k_y$=0 and $m_z$=2. The Weyl nodes are located at (0,0,$\pm\pi/2$). (d) Band structure on the $E$-$k_x$ plane with $k_y$=0, $k_z$=$\pi/2$ and $m_z$=2. The dispersion relation is quadratic in the $k_x$ direction near the Weyl node. Other parameters are $t_x$=$t_y$=1.}
\end{figure}

We consider the tight-binding Hamiltonian that describes the DWSM on a simple cube lattice with the lattice constant $a\equiv1$. The Hamiltonian has the form of $H(\textbf{\textit{k}})=\sum_{\textbf{\textit{k}}}(a^{\dagger}_{{\textbf{\textit{k}}}\uparrow},a^{\dagger}_{{\textbf{\textit{k}}}\downarrow})h(\textbf{\textit{k}})\begin{pmatrix}
a_{{\textbf{\textit{k}}}\uparrow} \\
a_{{\textbf{\textit{k}}}\downarrow}
\end{pmatrix}$ with~\cite{MXY}
\begin{eqnarray}\label{H}
h(\textbf{\textit{k}})&=&{t_x}(\cos{k_x}-\cos{k_y}){\sigma_x}+{t_y}\sin{k_x}\sin{k_y}{\sigma_y}\nonumber\\ %
&\;&+({m_z}-\cos{k_x}-\cos{k_y}-{t_z}\cos{k_z}){\sigma_z},
\end{eqnarray}
where $t_{x,y,z}$ and $m_{z}$ are model parameters. $\sigma_{x,y,z}$ are Pauli matrices. The Hamiltonian (\ref{H}) has the form of $h(\textbf{\textit{k}})$=$\textbf{\textit{d}}(\textbf{\textit{k}})$$\cdot$$\boldsymbol{\sigma}$. Diagonalizing the Hamiltonian (\ref{H}) and we can get the energy spectrums $E_{k}$=$\pm\sqrt{d_{x}^{2}(\textbf{\textit{k}})+d_{y}^{2}(\textbf{\textit{k}})+d_{z}^{2}(\textbf{\textit{k}})}$. The conduction and valence bands touch each other when $d_{x}(\textbf{\textit{k}})$=$d_{y}(\textbf{\textit{k}})$=$d_{z}(\textbf{\textit{k}})$=0. By solving the equations, we can obtain a pair of twofold degenerate double Weyl points [see Figs.~\ref{figure1}(c) and \ref{figure1}(d)] located at (0,0,$\pm\arccos$(($m_{z}-$2)/$t_{z}$) for 2$-t_z$$<$$m_{z}$$<$2+$t_{z}$. When $t_{z}-$2$<$$m_{z}$$<$2$-t_{z}$, the energy spectrums open a topological nontrivial gap and this model corresponds to the 3D QAHI phase [see Fig.~\ref{figure1}(b)]. When $m_{z}$$>$$t_{z}$+2, the spectrums open a gap as well but the model now is topological trivial and generates a NI phase. According to the above inequality equations, we draw the phase diagram in the clean limit in Fig.~\ref{figure1}(a). In this work, we choose $t_{z}$=0.5 and $t_{x}$=$t_{y}$=1.

It's not convenient to include the disorder effect into Eq.~(\ref{H}) in $k$-space, so we take the Fourier transform
$a_{\textbf{\textit{k}}\sigma}$=(1/$\sqrt{V}$)$\sum_{\textbf{\textit{r}}\sigma}e^{i\textbf{\textit{k}}\textbf{\textit{r}}}a_{\textbf{\textit{r}}\sigma}$ and get the real space Hamiltonian
\begin{eqnarray}\label{H2}
H_{\rm real}&=&\sum\limits_{\textbf{\textit{r}}}{a^{\dagger}_{\textbf{\textit{r}}}}{T_0}{a_{\textbf{\textit{r}}}}+\big({a^{\dagger}_{\textbf{\textit{r}}+x}}T_{x}{a_{\textbf{\textit{r}}}}
+{a^{\dagger}_{\textbf{\textit{r}}+y}}T_{y}{a_{\textbf{\textit{r}}}}\nonumber+{a^{\dagger}_{\textbf{\textit{r}}+z}}T_{z}{a_{\textbf{\textit{r}}}}\\
&\;&+{a^{\dagger}_{\textbf{\textit{r}}+x+y}}T_{xy1}{a_{\textbf{\textit{r}}}}+{a^{\dagger}_{\textbf{\textit{r}}+x-y}}T_{xy2}{a_{\textbf{\textit{r}}}}+h.c.\big),
\end{eqnarray}
where ${a^{\dagger}_{\textbf{\textit{r}}}}=(a^{\dagger}_{{\textbf{\textit{r}}}\uparrow},a^{\dagger}_{{\textbf{\textit{r}}}\downarrow})$ with $a^{\dagger}_{\textbf{\textit{r}}\sigma}$ $(a_{\textbf{\textit{r}}\sigma})$ being the creation(annihilation) operator at site \textbf{\textit{r}} for the electron with spin $\sigma$=$\uparrow$,$\downarrow$, and \textit{x,y,z} denote the hopping directions. $T_{0}$ is the on-site energy, $T_{x}$, $T_{y}$ and $T_{z}$ are the nearest-neighbor hopping matrices along the \textit{x,y,z} axis. $T_{xy1}$ ($T_{xy2}$) is the next-nearest-neighbor hopping matrix along the $x$$+y$ ($x$$-y$) direction. Here $T_{0}$=$m_{z}\sigma_{z}$, $T_{x}$=$\frac{1}{2}$($\sigma_{x}$$-\sigma_{z}$), $T_{y}$=$-\frac{1}{2}$($\sigma_{x}$$+\sigma_{z}$), $T_{z}$=$-\frac{1}{2}t_{z}{\sigma_{z}}$, $T_{xy1}$=$-\frac{1}{4}{\sigma_{y}}$ and $T_{xy2}$=$\frac{1}{4}{\sigma_{y}}$.

In this work, the disorder effect is introduced by adding the random variable $H_{\rm disorder}$~\cite{CR3} to Eq.~(\ref{H2}),
\begin{eqnarray}
H_{\rm disorder}&=&\sum_\textbf{\textit{r}}{a^{\dagger}_{\textbf{\textit{r}}}}\big[U_x\sigma_x+U_y\sigma_y+U_z\sigma_z+U_0\sigma_0\big]{a_{\textbf{\textit{r}}}}.
\end{eqnarray}
The first three terms denote the orbital disorder and the last term is the on-site nonmagnetic disorder. $U_{x,y,z,0}$ are uniformly distributed within [-$W_{x,y,z,0}$/2,$W_{x,y,z,0}$/2] with $W_{x,y,z,0}$ representing the disorder strength.

We use the transfer matrix method to numerically calculate the localization length to determine the phase boundary induced by disorder~\cite{Milde,MacKinnon,MacKinnon2,Kramer,Slevin}. We consider a quasi-one-dimensional long bar with the system volume $L_{x}$$\times$$L_{y}$$\times$$L_{z}$ and apply periodic boundary conditions in the $x$ and $y$ directions. We divide the system into slices with each slice's cross-section $L_{x}$$\times$$L_{y}$=$L$$\times$$L$. We let $\psi_{n}$ and $H_{n,n}$ be the wave function and hopping matrix of the $n$th slice. $H_{n,n-1}~(H_{n.n+1})$ is the hopping matrix between the $n$th and $n$-1($n$+1)th slice. The Schr\"{o}dinger equation can be written as~\cite{Milde,CR3}
\begin{eqnarray}\label{SEQ}
H_{n,n}\psi_n+H_{n,n+1}\psi_{n+1}+H_{n,n-1}\psi_{n-1}=E\psi_n.
\end{eqnarray}
Eq.~(\ref{SEQ}) can be further expressed as
\begin{eqnarray}
\begin{pmatrix}
\psi_{n+1} \\
\psi_{n}
\end{pmatrix}
=
T_n
\begin{pmatrix}
\psi_{n} \\
\psi_{n-1}
\end{pmatrix},
\end{eqnarray}
with the transfer matrix
\begin{eqnarray}
T_{n}=\begin{pmatrix}
      H^{-1}_{n,n-1}(E-H_{n,n}) & ~~~-H^{-1}_{n,n+1}H_{n,n-1} \\
      1 & 0
      \end{pmatrix}.
\end{eqnarray}
$Q_{L_{z}}$=$\prod^{L_{z}}_{n=1}T_{n}$ is the product of transfer matrix and there exists a limit matrix $\Gamma$=$\lim\limits_{L_{z} \to +\infty}({Q_{L_{z}}^\dagger}{Q_{L_{z}}})^{\frac{1}{2L_{z}}}$. By diagonalizing the matrix $\Gamma$, we obtain the normalized eigenvectors $\{u_{i}\}$. The Lyapunov exponent is defined as
\begin{eqnarray}
\gamma_{i}=\frac{1}{L_z}\lim\limits_{L_{z} \to +\infty}\ln\big\|Q_{L_z}u_i\big\|.
\end{eqnarray}
The localization length $\lambda$ characterizes the largest possible extension of a state and is defined as the inverse of the smallest positive $\gamma_i$ by $\lambda$=$1/\gamma_{min}$. Besides, to determine the critical point of the phase transition, what we need is the normalized localization length $\Lambda$ defined as $\Lambda$=$\lambda/L$, where $L$ is the side length of each slice. In the following calculation, we choose $L$=8, 10, 12 and 14 ignoring the unit since we fix the lattice constant to be a$\equiv$1. In general, in metallic phase, the rate of change of normalized localization length $\Lambda$ with the system size $L$ satisfies $d\Lambda/dL$$>$0. And in the insulator phase, $d\Lambda/dL$$<$0. A phase transition happens when $d\Lambda/dL$=0.

We can view the DWSM as coupled multiple 2D systems labeled by $k_z$ as $H_{0}(k_z)$ with the Weyl nodes located at $(0,0,\pm{k_0})$. For $k_z$$\in$($-k_0$,$k_0$), the system is a 2D Chern insulator and contributes a quantized Hall conductivity $\sigma_{xy}^{2D}$=$2e^2$/$h$. For $k_z$$\notin$($-k_0$,$k_0$), the system is topological trivial and therefore has no contribution to the Hall conductivity. So the total Hall conductivity of the 3D system is $\sigma_{xy}^{3D}$=$\sum_{kz}\sigma_{xy}^{2D}(k_z)/L_z$. In the presence of disorder, the Hall conductivity can be calculated by the noncommutative geometry method~\cite{Prodan,Prodan2,Prodan3}, which is used to calculate the Chern number in real space now that the disorder can break the translation symmetry in $k$-space. The Chern number in real space can be expressed as
\begin{eqnarray}\label{Chern}
C=-\frac{2\pi{i}}{N^2}\sum\limits_{n,\alpha}\langle{n,\alpha}|P[-i[x_1,P],-i[x_2,P]]|{n,\alpha}\rangle,
\end{eqnarray}
where
\begin{eqnarray}\label{duiyi}
[x_i,P]=i\sum\limits_{m=1}^Q{c_m}(e^{-im\textbf{\textit{x}}\Delta_i}Pe^{im\textbf{\textit{x}}\Delta_i}-e^{im\textbf{\textit{x}}\Delta_i}Pe^{-im\textbf{\textit{x}}\Delta_i}).
\end{eqnarray}
In Eqs.~(\ref{Chern}) and (\ref{duiyi}), $N$ is the sample size along the $x$ and $y$ directions, $\Delta_i$=$2\pi$/$N$. $P$ is the projection operator of the occupied state in real space and $|n,\alpha\rangle$ is the real space coordinate. $Q$ takes the integer between $0$ and $N$/2. The corresponding Chern number will be more accurate for larger $Q$. Here we set $Q$=8.

\section{NUMERICAL RESULTS}\label{sec:Results}
In this section, we numerically calculate the localization length and the Hall conductivity based on the above methods and draw the global phase diagrams in the presence of disorder. In Sec.\ref{sec:CPD} and Sec.\ref{sec:IOD}, we consider the nonmagnetic disorder and the $\sigma_{x}$ orbital disorder, respectively. And in Sec.\ref{sec:IID}, we investigate the combined effect of three sorts of orbital disorders.
\subsection{ON-SITE NONMAGNETIC DISORDER}\label{sec:CPD}
\begin{figure}[htbp]
  \centering
  \includegraphics[scale=0.6]{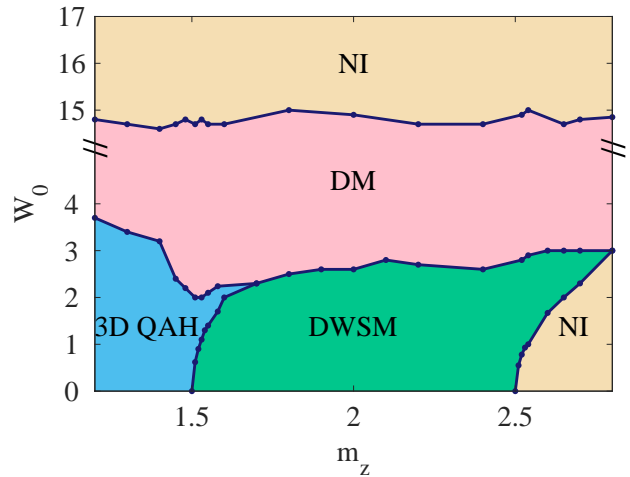}
  \caption{\label{figure2}Phase diagram on the $W_0$-$m_z$ plane for $t_z$=0.5 and $t_x$=$t_y$=1. The dark blue solid lines with dots are identified by the transfer matrix method. The accurate phases are determined by the Hall conductivity $\sigma_{xy}$.}
\end{figure}

\begin{figure}[htbp]
  \centering
  \includegraphics[scale=0.24]{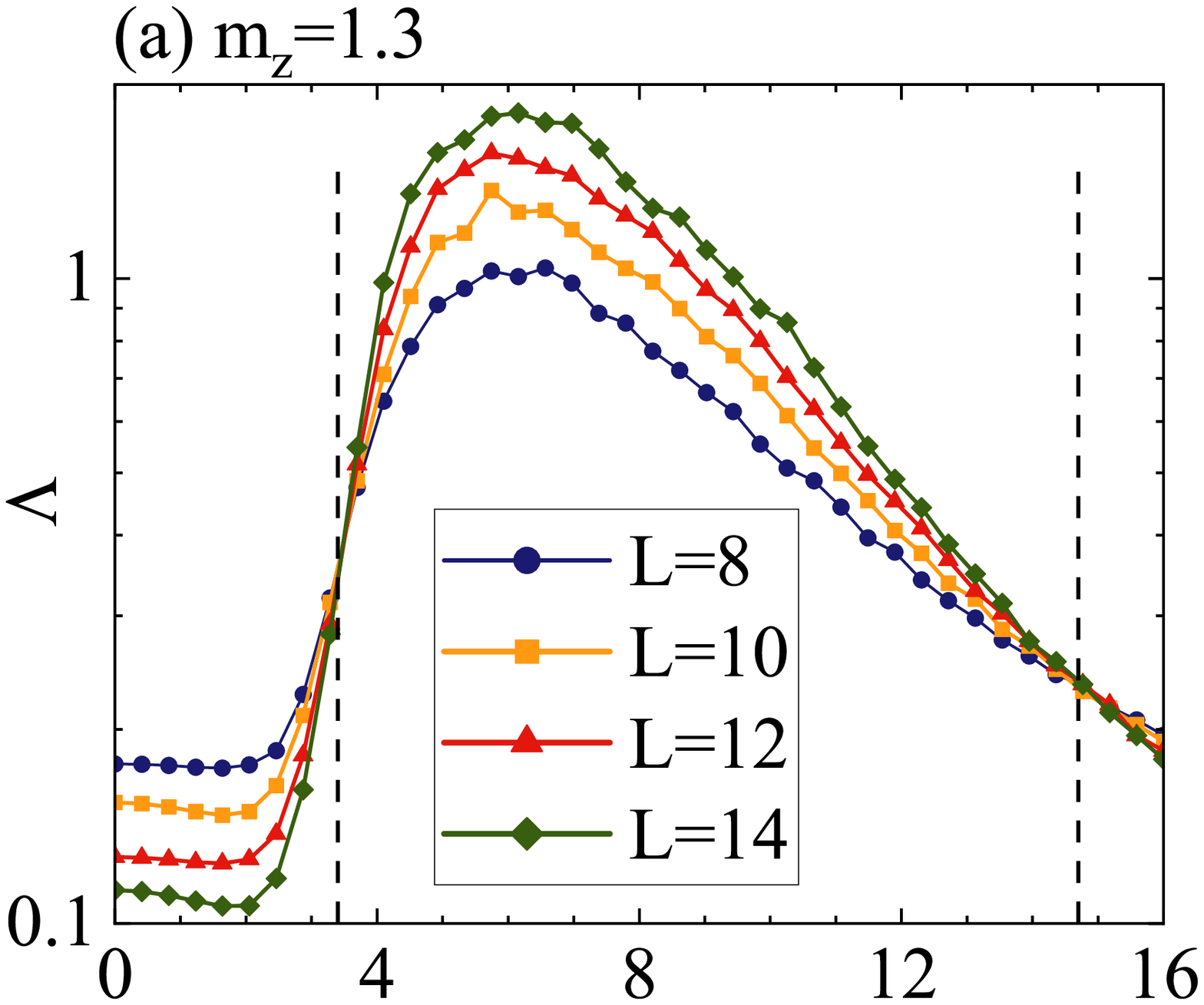}
  \includegraphics[scale=0.24]{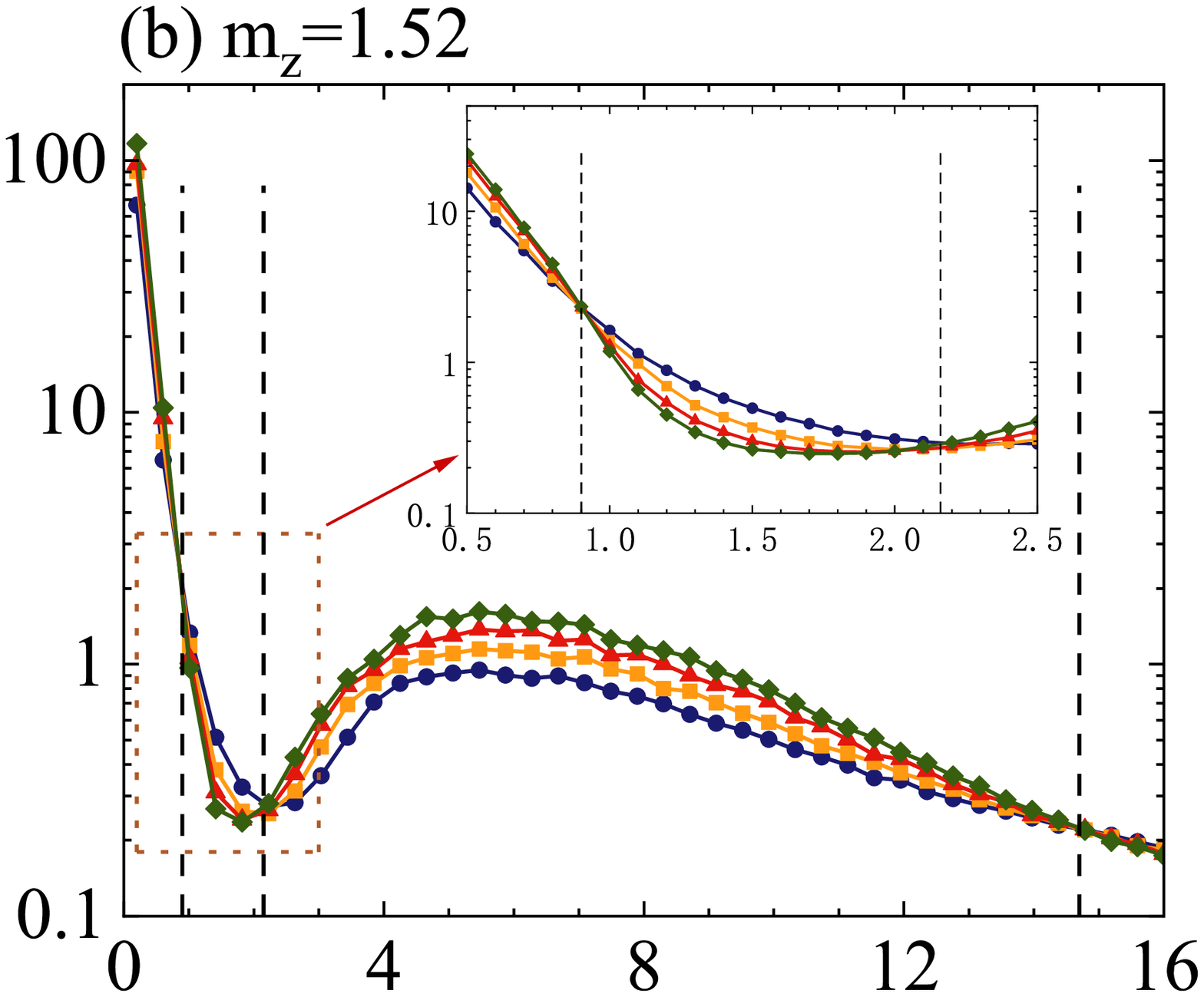}

  \includegraphics[scale=0.24]{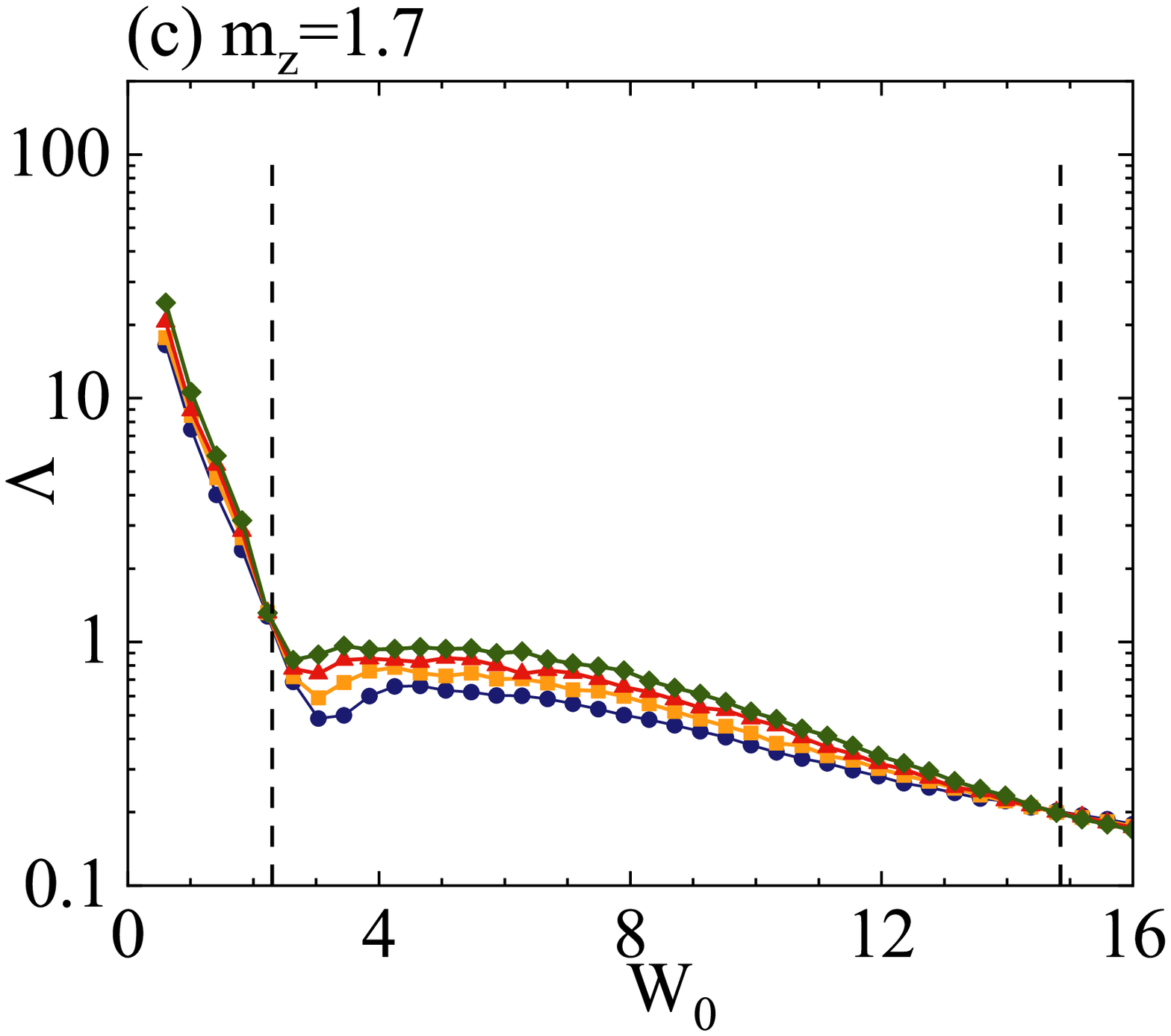}
  \includegraphics[scale=0.24]{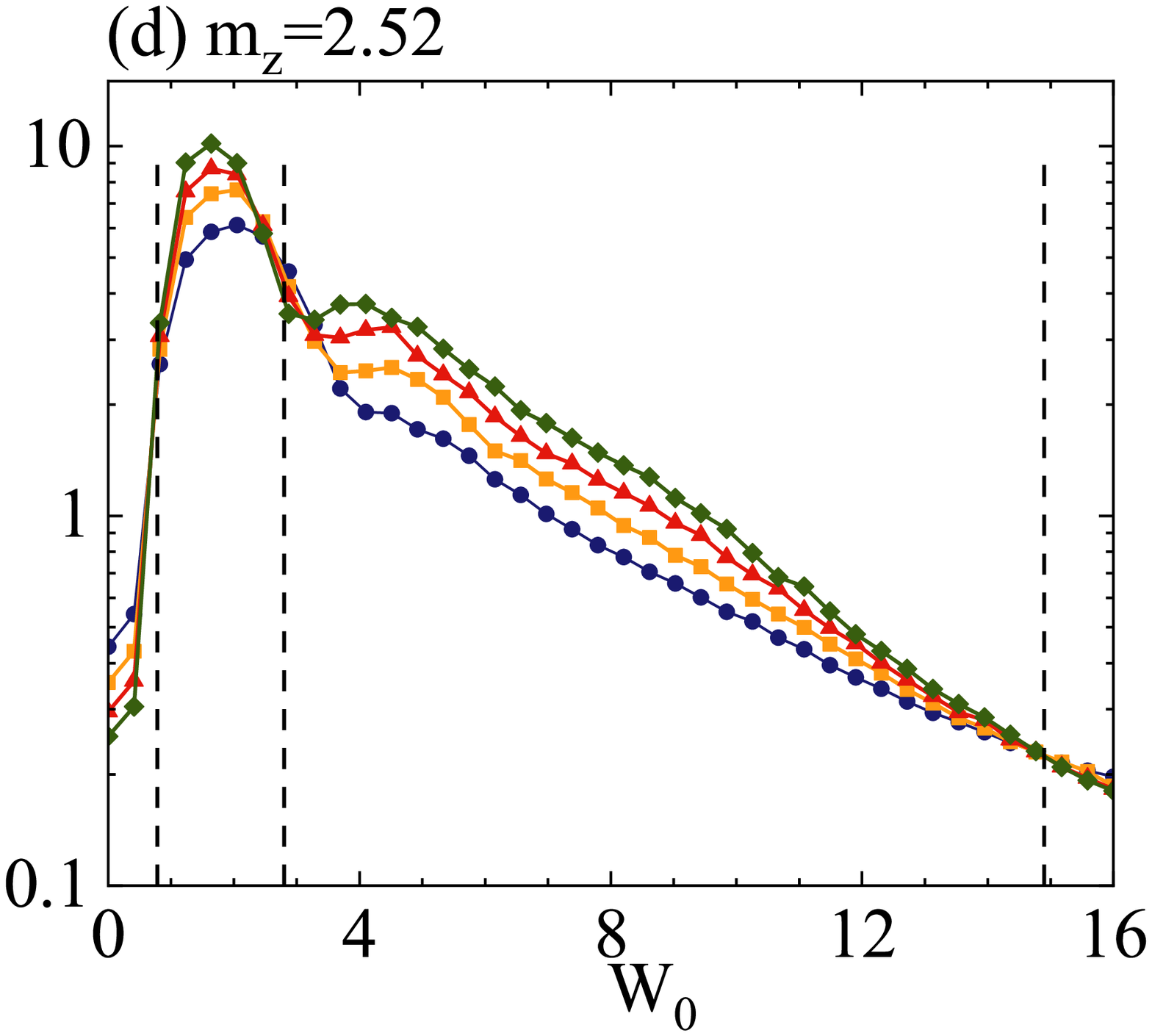}
  \caption{\label{figure3} The normalized localization length $\Lambda$ is plotted as a function of the disorder strength $W_0$ for the values of (a) $m_z$=1.3, (b) 1.52, (c) 1.7 and (d) 2.52. The inset in Fig.~\ref{figure3}(b) shows the details near the phase transition points. Four colored curves associates with different side lengths of the cross section. Other parameters are $t_z$=0.5 and $t_x$=$t_y$=1.}
\end{figure}
\begin{figure}
  \centering
  \includegraphics[scale=0.22]{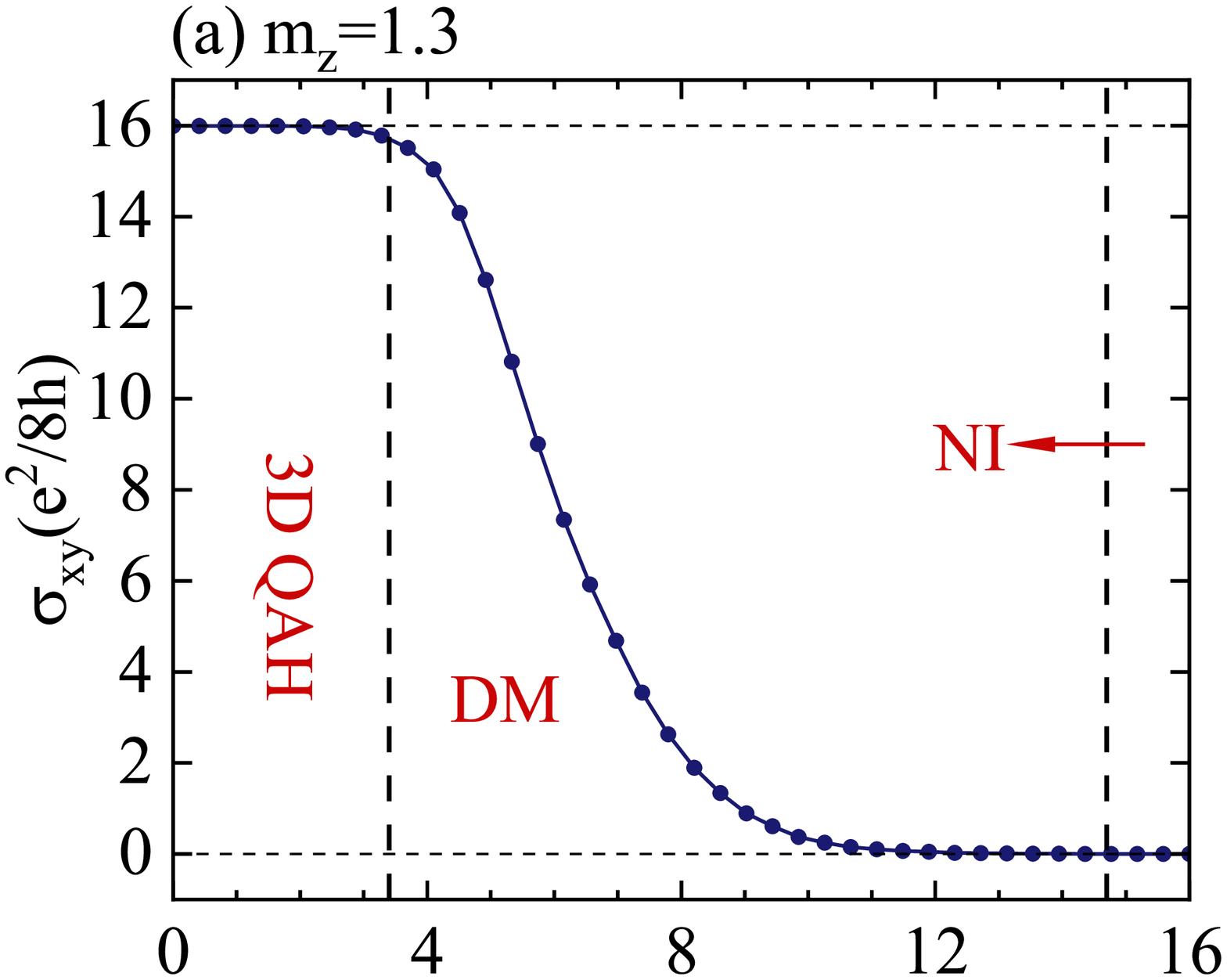}
  \includegraphics[scale=0.22]{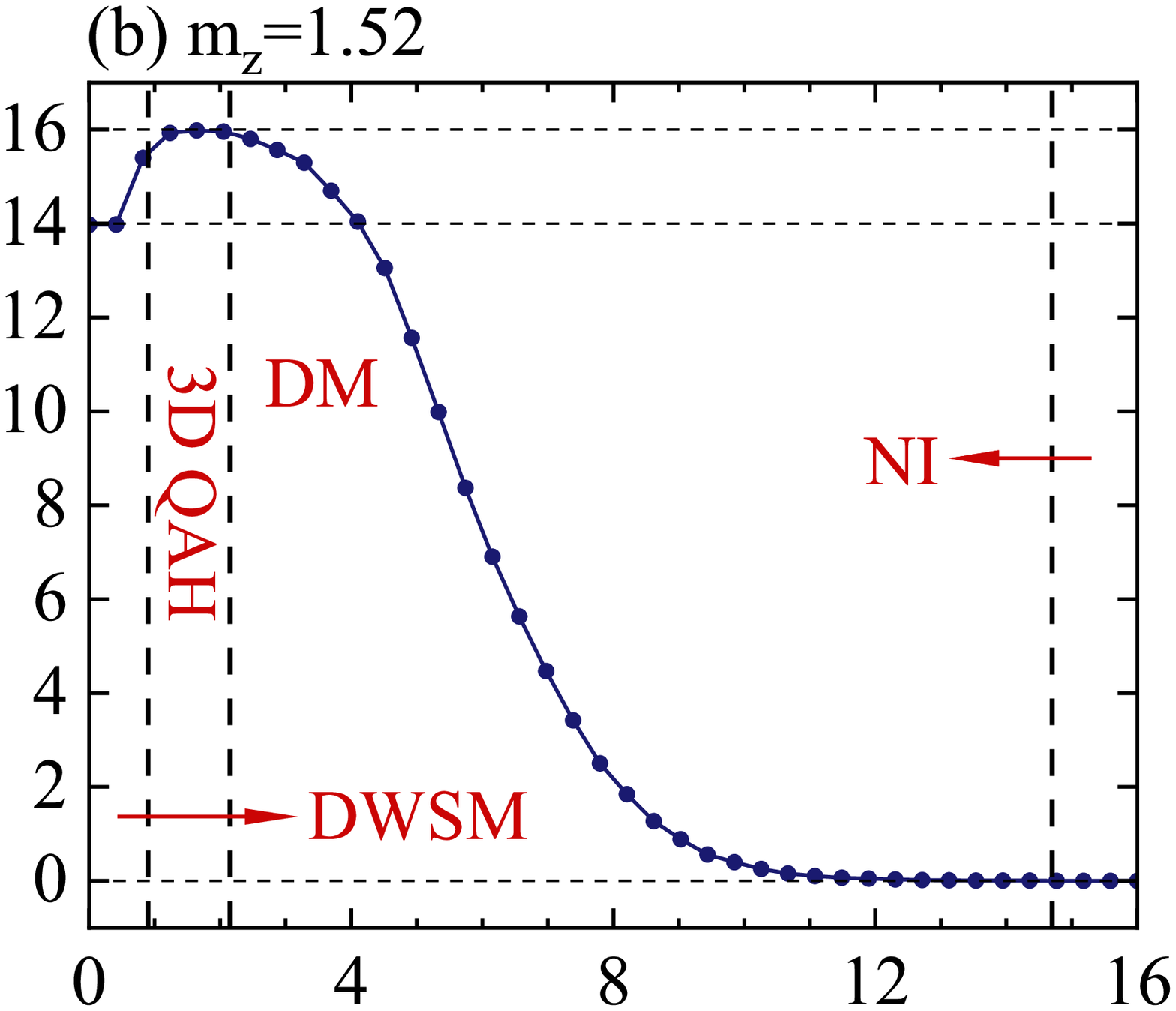}

  \includegraphics[scale=0.22]{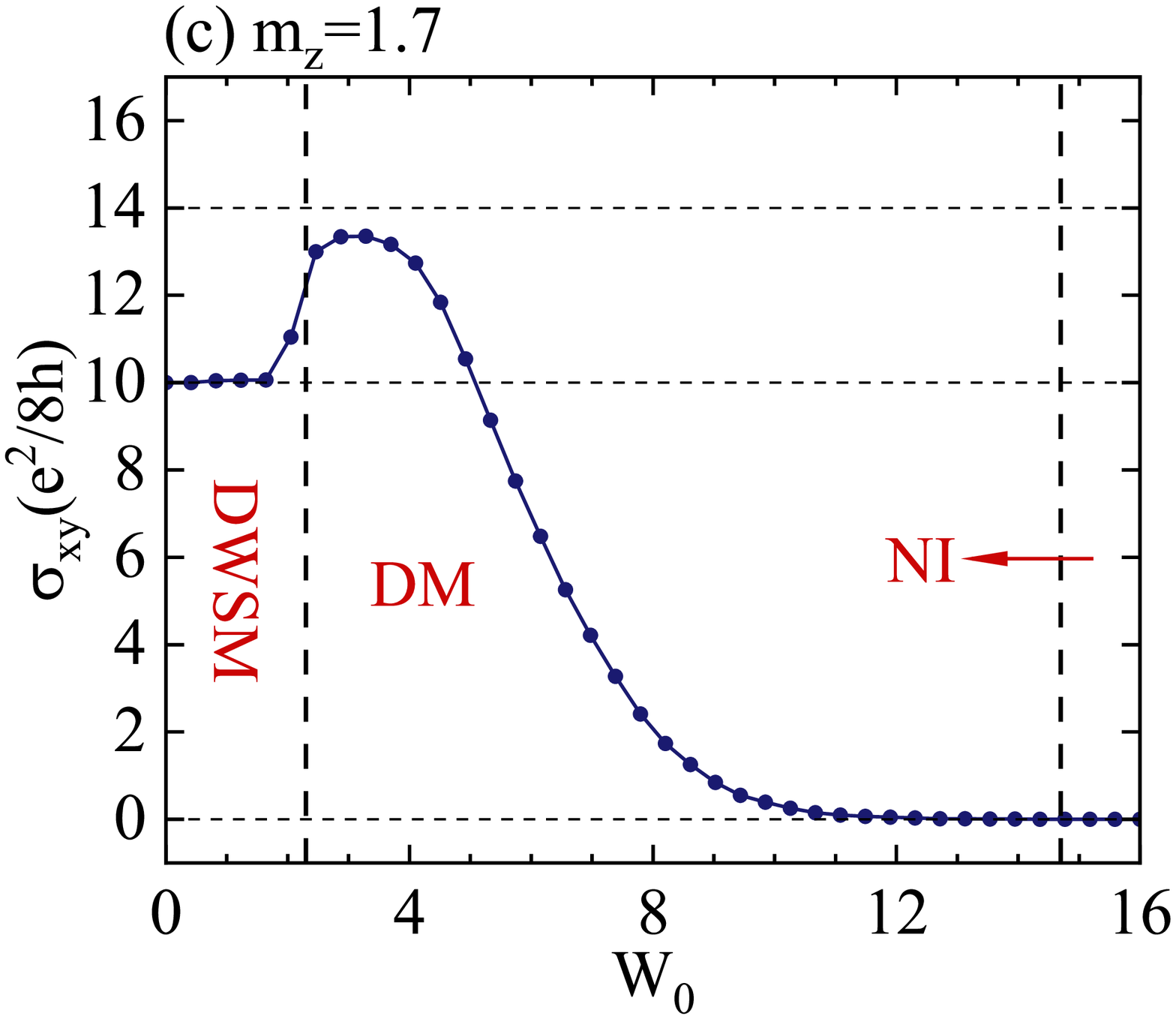}
  \includegraphics[scale=0.22]{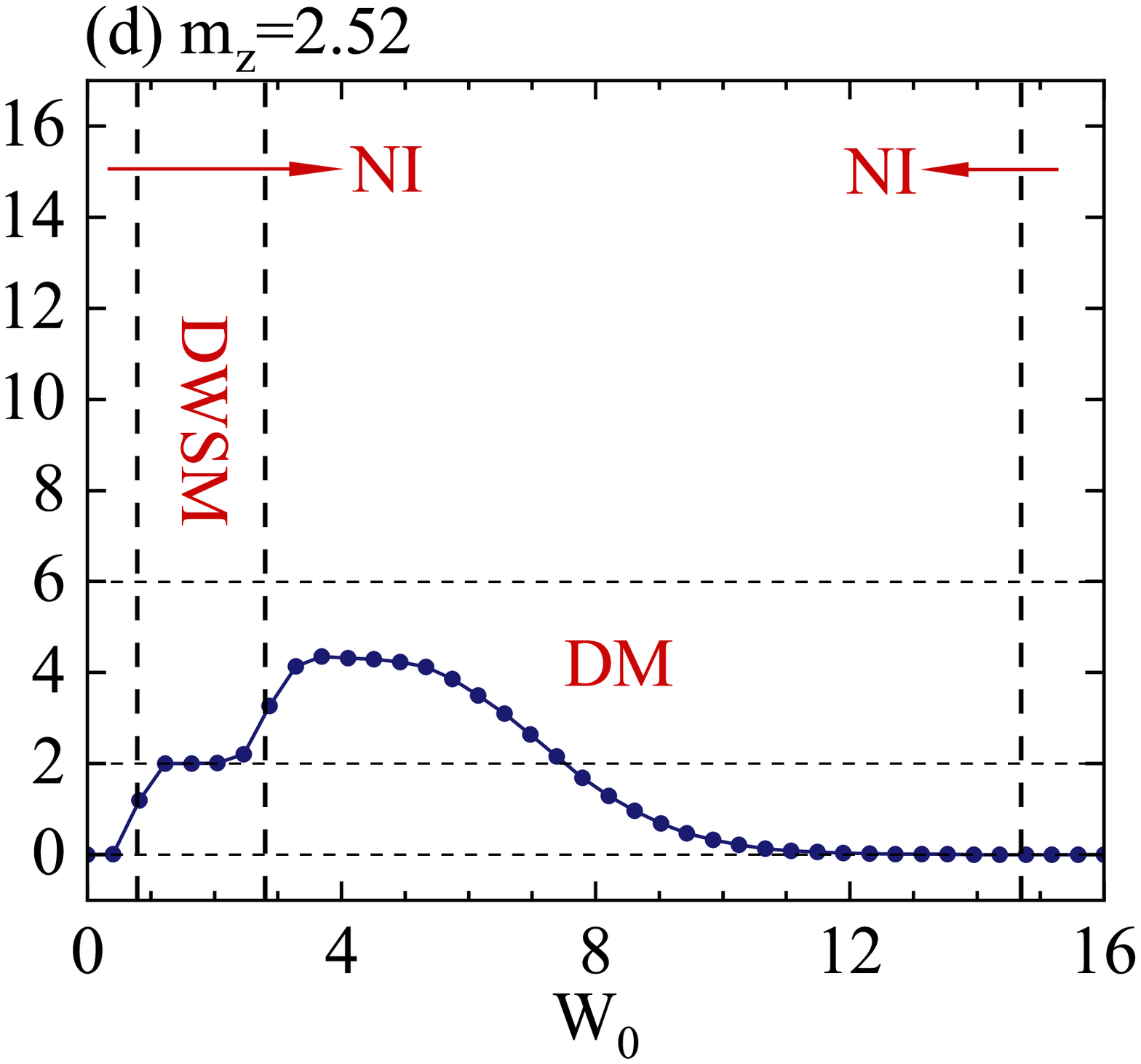}
  \caption{\label{figure4}The hall conductivity $\sigma_{xy}$ as a function of the disorder strength $W_0$ for the values of (a) $m_z$=1.3, (b) 1.52, (c) 1.7 and (d) 2.52. The system size is 40$\times$40$\times$8. Other parameters are $t_z$=0.5 and $t_x$=$t_y$=1.}
\end{figure}

By adjusting the parameter $m_z$ in Eq.~(\ref{H}), the system in the clean limit can be 3D QAH, DWSM or NI phase [marked by the black dots in Fig.~\ref{figure1}(a)] with $t_z$=0.5 and $t_{x,y}$=1. Here we set $m_z$$\in$[1.2,2.8] and investigate the phase transition caused by the nonmagnetic disorder. We calculate the renormalized localization length to determine the phase boundary and the Hall conductivity to distinguish different phases. The 3D QAH phase has a quantized Hall conductivity. In DWSM phase, the Hall conductivity is fractional quantized because of the finite Fermi arc between two Weyl nodes. The global phase diagram is summarized in Fig.~\ref{figure2}. We pick four typical results of the localization length and the Hall conductivity for $m_z$=1.3, 1.52, 1.7 and 2.52, and plot them in Fig.~\ref{figure3} and Fig.~\ref{figure4}.

For $m_z$=1.3, the rate of change $d\Lambda$/$dL$ of the normalized localization length is a negative value, i.e., the $\Lambda$ decreases with the increase of side length $L$ from 8 to 12 for a fixed $W_0$$<$3.4. Meanwhile, the Hall conductivity stays quantized $\sigma_{xy}$=$16e^2$/$8h$, which indicates that the 3D QAH phase is robust against the weak disorder. In Fig.~\ref{figure3}(a), $d\Lambda$/$dL$=0 at $W_0$=3.4 shows the existence of a phase transition. By increasing the disorder strength, $d\Lambda$/$dL$ turns to a positive value and the Hall conductivity $\sigma_{xy}$ becomes non-quantized, thus the system evolves into the DM phase. Further increasing the disorder strength, $d\Lambda/dL$ again becomes the negative value and the second phase transition happens at $W_0$=14.7. The vanishing Hall conductivity $\sigma_{xy}$=0 in Fig.~\ref{figure4}(a) manifests that the system is localized by the strong disorder and evolves into the NI phase.

\begin{figure}
  \centering
  \includegraphics[scale=0.36,trim=165 65 265 105,clip]{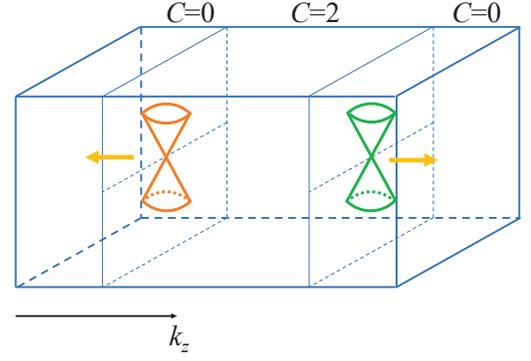}
  \caption{\label{weylcone}The schematic diagram of the Weyl cone in the Brillouin zone. The direction of arrow represents the direction of motion of the Weyl node. The difference of the color for the Weyl cones refers to different chirality.}
\end{figure}

For $m_z$=1.52, a phase transition from DWSM phase to 3D QAH phase occurs as evidenced by the Hall conductivity transition from $14e^2$/$8h$ to $16e^2$/$8h$ in Fig.~\ref{figure4}(b) and $d\Lambda$/$dL$ changing from a positive to negative value in Fig.~\ref{figure3}(b) at $W_0$=0.9. This phenomenon can be explained by SCBA [see Sec.\ref{sec:SCBA}]. Based on SCBA, the nonmagnetic disorder renormalizes the mass term $m_z$ in the Hamiltonian~(\ref{H}). And $m_z$ determines the positions of the Weyl nodes in the Brillouin zone. According to Eq.~(\ref{mz}), the renormalized mass term $\tilde{m}_z$ decreases with the increase of the disorder strength $W_0$, which causes a pair of Weyl nodes depart from each other and move to the boundary of the Brillouin zone [see Fig.~\ref{weylcone}]. When they arrive at the zone boundary, they annihilate pairwise and a nontrivial bulk gap arises, and the system realizes the phase transition from DWSM to 3D QAH phase. Further increasing disorder strength to $W_0$=2.16, $d\Lambda$/$dL$ turns to a positive value and the Hall conductivity becomes non-quantized, which means the system evolve into the DM phase. Likewise, the system is localized when the disorder strength surpasses the value of $W_0$=14.8.

For $m_z$=1.7, as shown in Figs.~\ref{figure3}(c) and \ref{figure4}(c), a DWSM-DM phase transition happens at $W_0$=2.3 as the disorder strength increases. In this parameter configuration, the 3D QAH phase disappears because the system has already entered the DM phase before the Weyl nodes reach the zone boundary. And for $m_z$=2.52, with increasing of the disorder strength, as shown in Figs.~\ref{figure3}(d) and \ref{figure4}(d), a phase transition takes place at $W_0$=0.78 and the Hall conductivity increases from $\sigma_{xy}$=0 to a fractional quantized value of $\sigma_{xy}$=$2e^2$/$8h$. This interesting phase transition can also be understood by SCBA. According to Eq.~(\ref{mz}), the renormalized mass term $\tilde{m}_z$ decreases as $W_0$ increases. When $\tilde{m}_z$ reaches the critical point corresponding to $m_z$=2.5 in Fig.~\ref{figure1}(a), this trivial bulk gap closes and a pair of Weyl nodes emerge at the center of the Brillouin zone, realizing the phase transition from NI phase to DWSM phase. With further increasing of the disorder strength, the system evolves from DWSM phase to DM phase, and finally recovers NI phase. The global phase diagram in Fig.~\ref{figure2} is obtained by repeatedly calculating the normalized localization length and the Hall conductivity at various values of $W_0$ and $m_z$.

\subsection{$\boldsymbol{\sigma_{x}}$ ORBITAL DISORDER}\label{sec:IOD}
\begin{figure}
  \centering
  \includegraphics[scale=0.6]{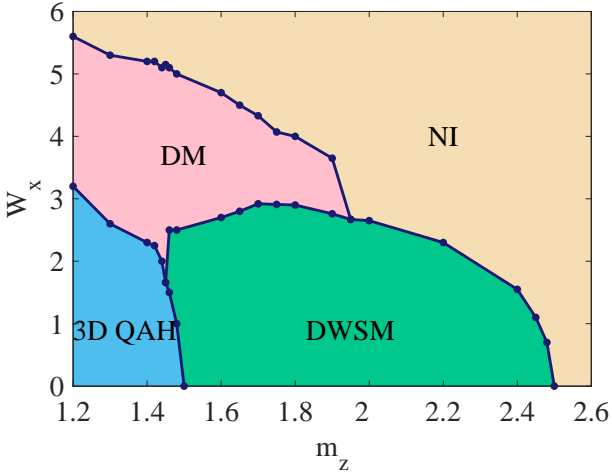}
  \caption{\label{figure5}Phase diagram on the $W_x$-$m_z$ plane for $t_z$=0.5 and $t_x$=$t_y$=1. The dark blue solid lines with dots are identified by the transfer matrix method. The accurate phases are determined by the Hall conductivity $\sigma_{xy}$.}
\end{figure}

\begin{figure}
  \centering
  \includegraphics[scale=0.24]{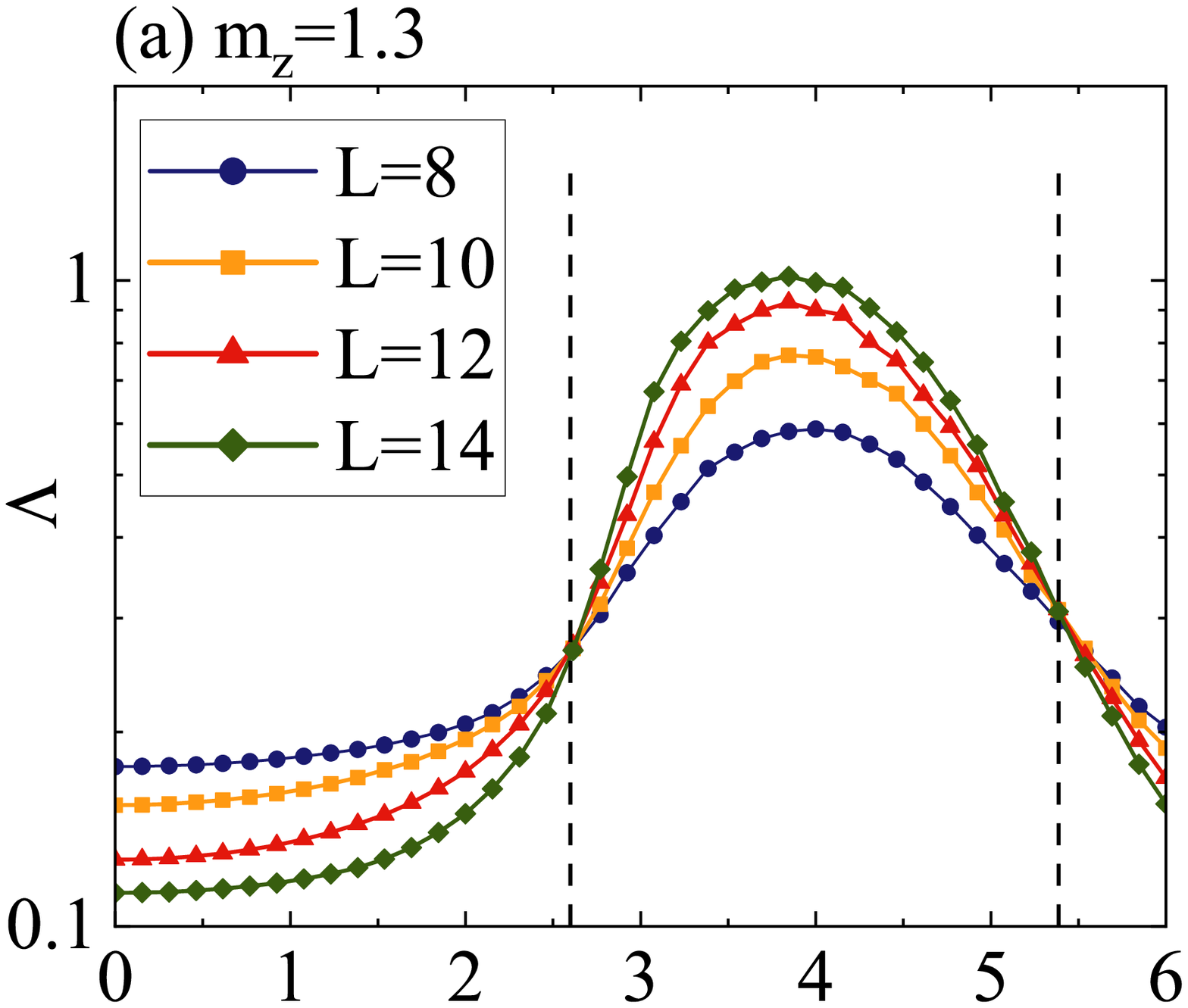}
  \includegraphics[scale=0.24]{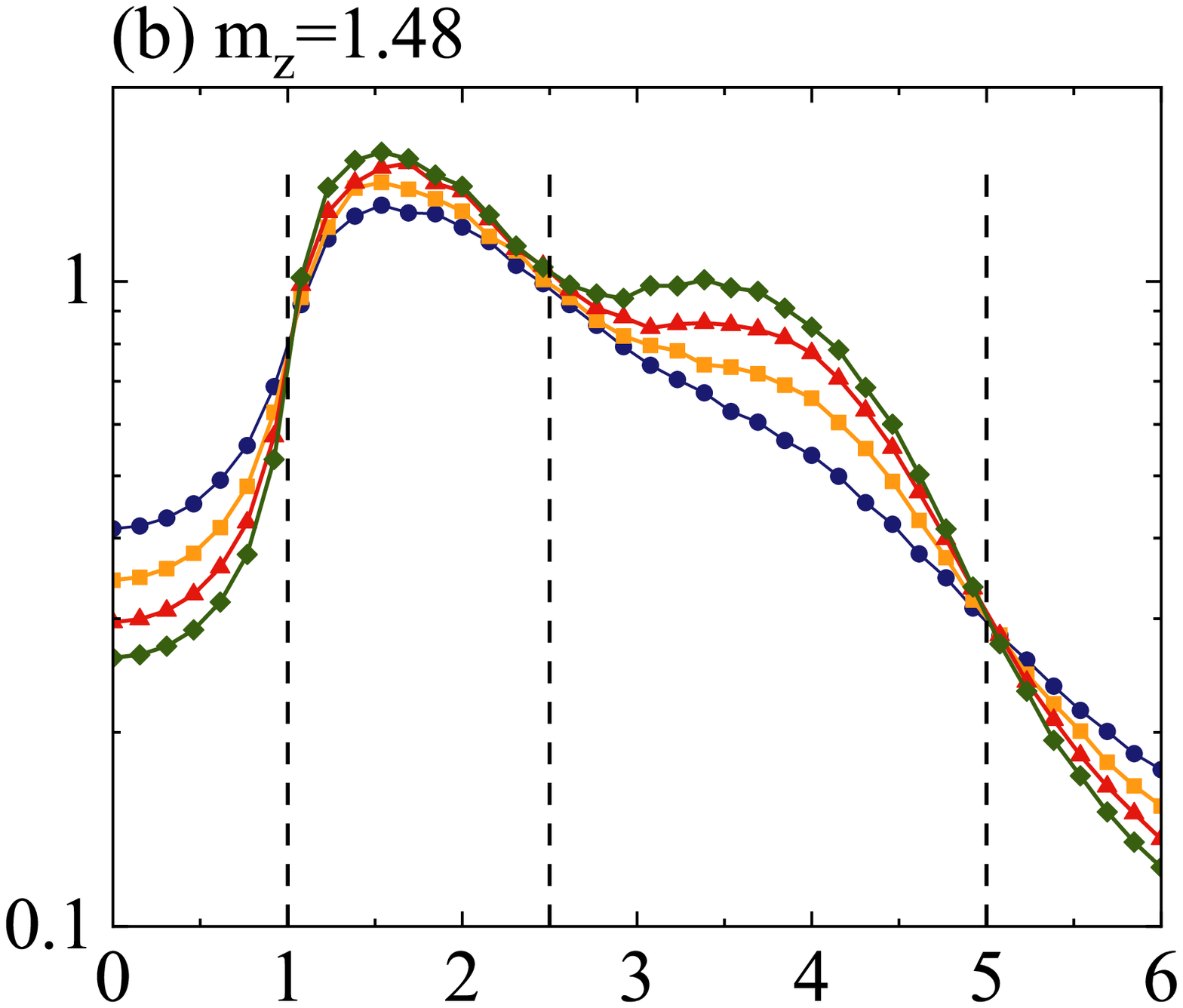}

  \includegraphics[scale=0.24]{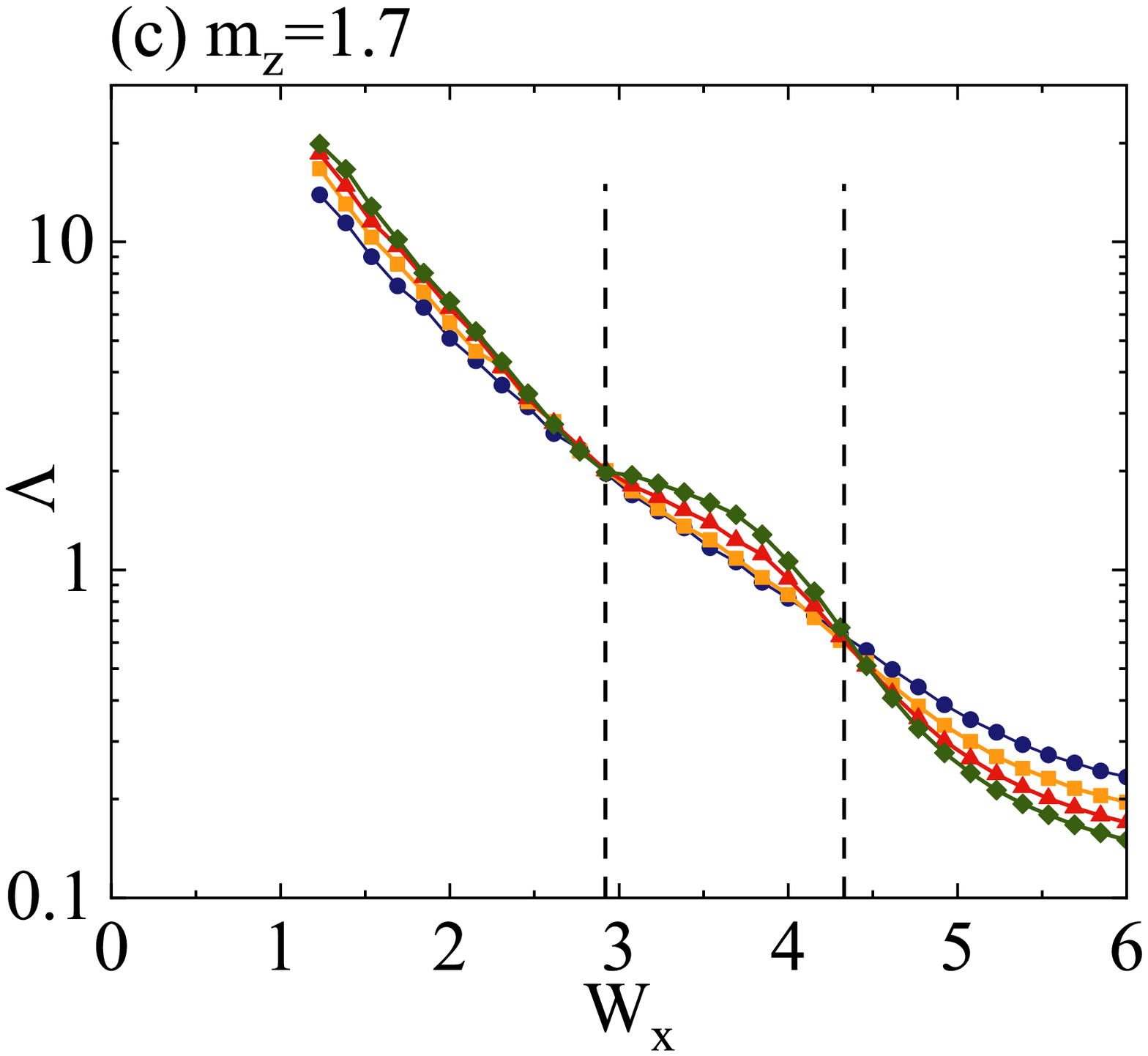}
  \includegraphics[scale=0.24]{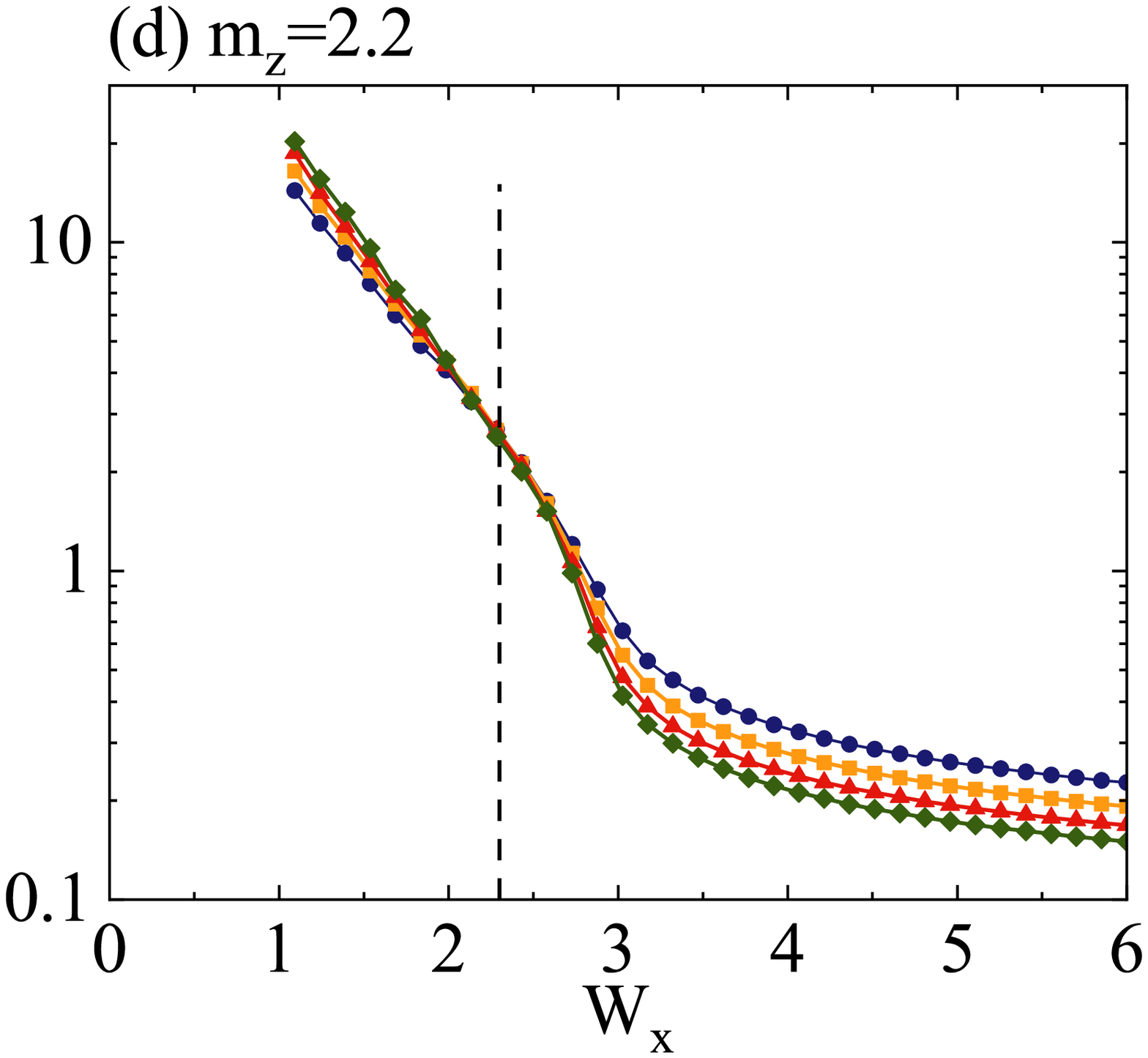}
  \caption{\label{figure6}The normalized localization length $\Lambda$ is plotted as a function of the disorder strength $W_x$ for the values of (a) $m_z$=1.3, (b) 1.48, (c) 1.7 and (d) 2.2. Four colored curves associates with different side lengths of the cross section. Other parameters are $t_z$=0.5 and $t_x$=$t_y$=1.}
\end{figure}

\begin{figure}
  \centering
  \includegraphics[scale=0.22]{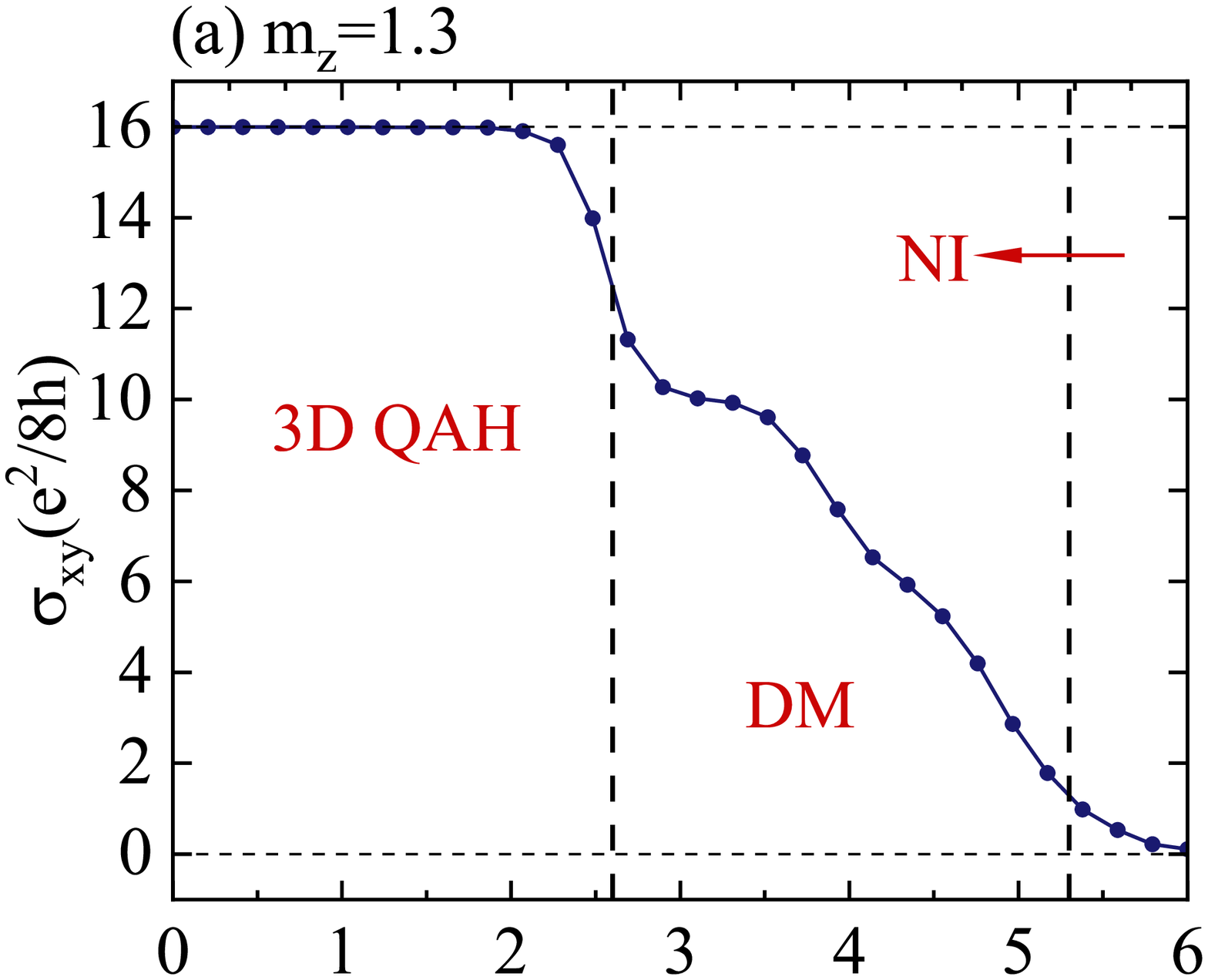}
  \includegraphics[scale=0.22]{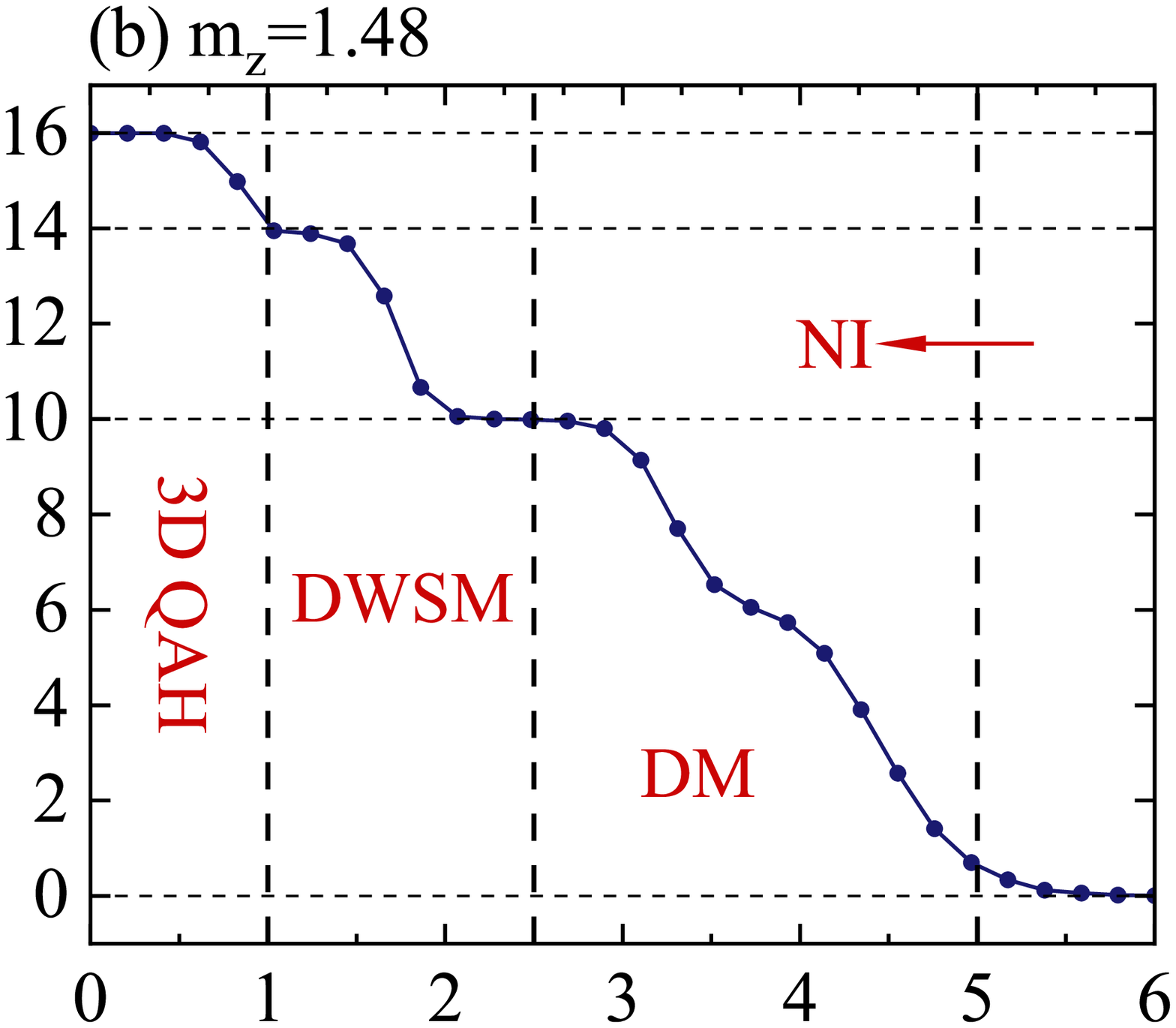}

  \includegraphics[scale=0.22]{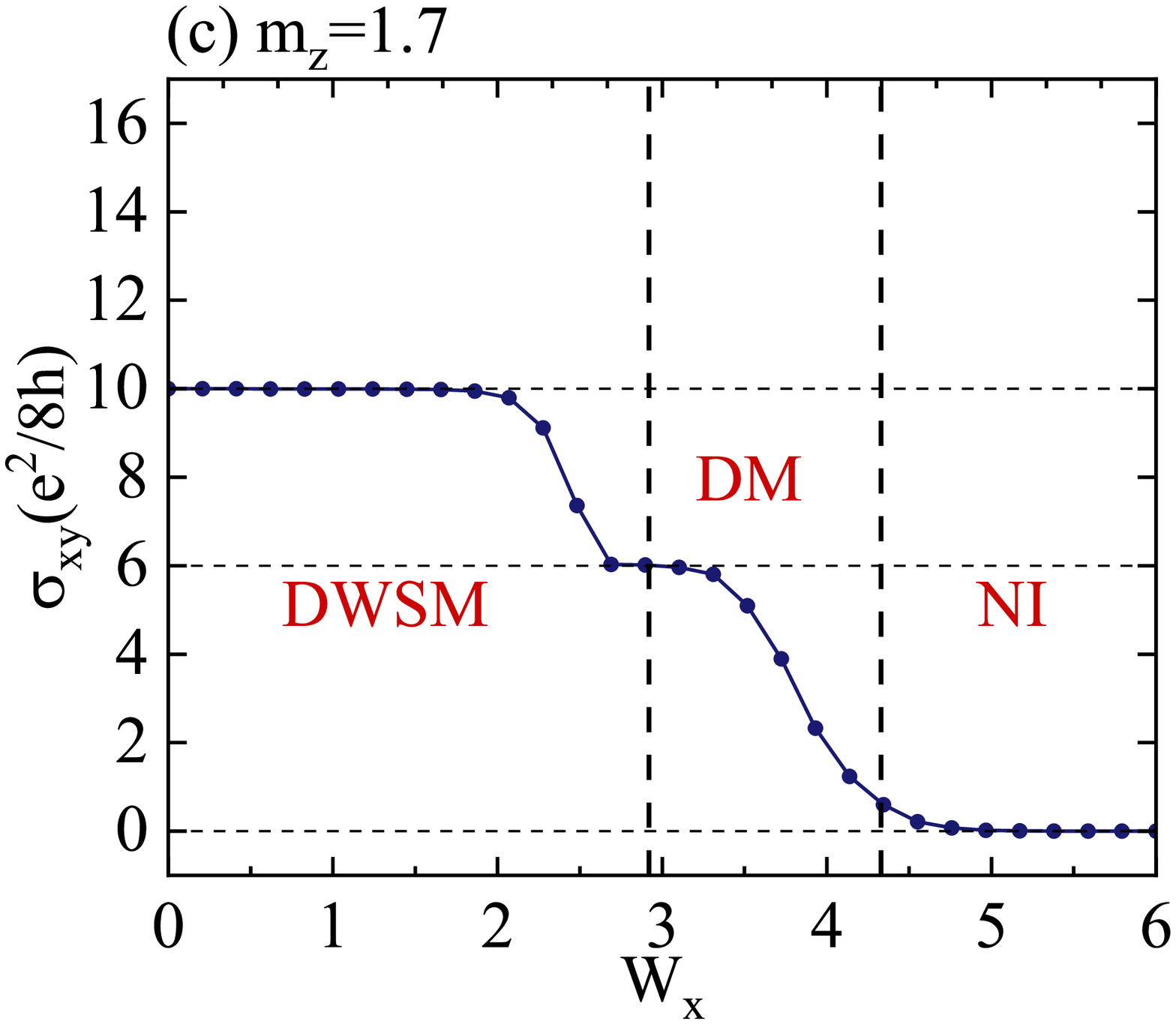}
  \includegraphics[scale=0.22]{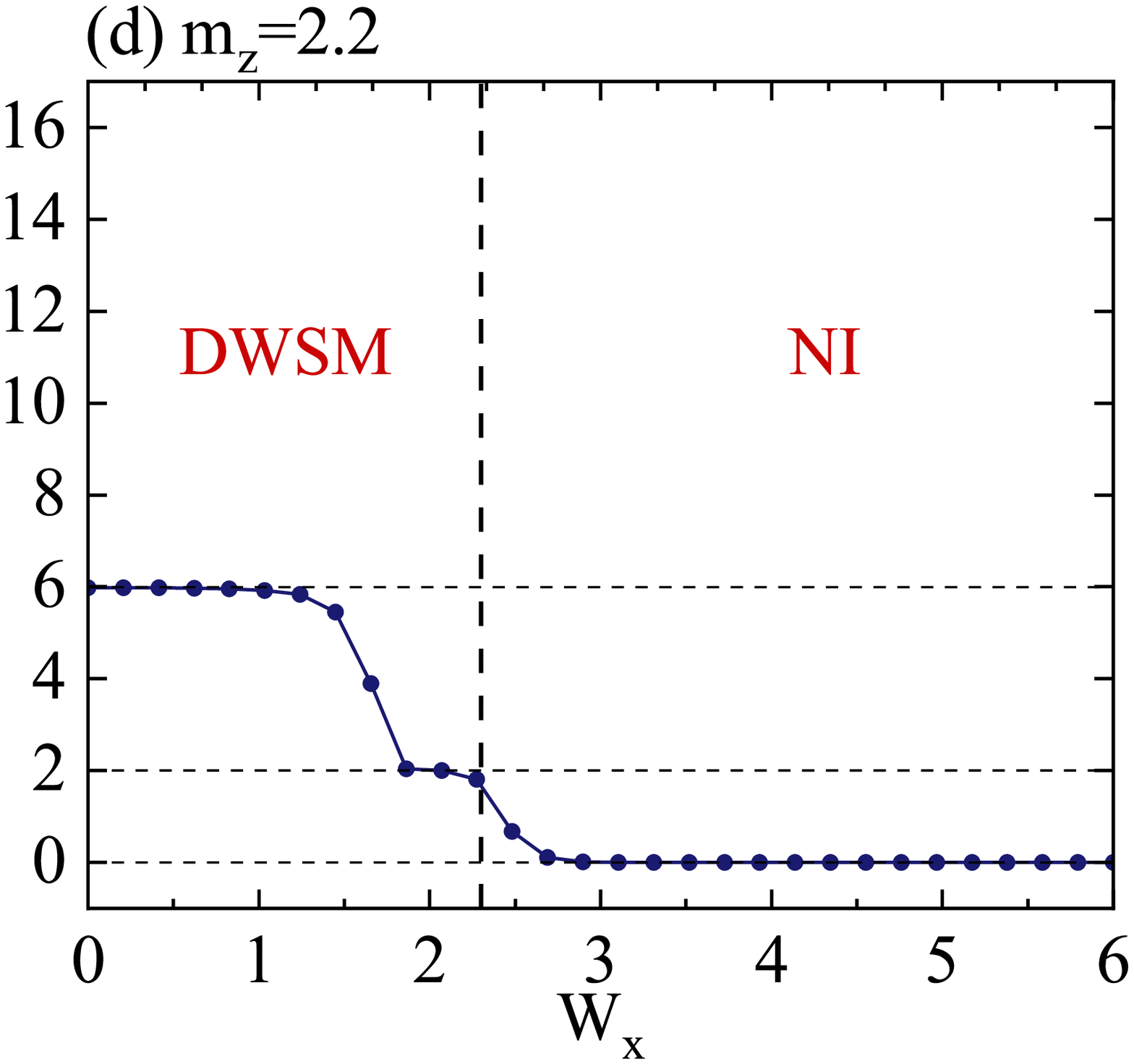}
  \caption{\label{figure7}The hall conductivity $\sigma_{xy}$ is plotted as a function of the disorder strength $W_x$ for the values of (a) $m_z$=1.3, (b) 1.48, (c) 1.7 and (d) 2.2. The system size is 40$\times$40$\times$8. Other parameters are $t_z$=0.5 and $t_x$=$t_y$=1.}
\end{figure}

There is another kind of disorder, i.e. the orbital disorder. The orbital disorder means that the spin-up electron could be randomly scattered to the spin-down state, thus is also named as the spin-flip disorder~\cite{QZH}. This type of disorder, which has been investigated in previous researches~\cite{CR3,SJT,HHH,QZH,HLH,YH}, is missing in DWSM. Thus we next study the phase transition generated by $\sigma_x$ orbital disorder. Fig.~\ref{figure5} shows the global phase diagram for different $m_z$ and disorder strengths $W_x$. Figs.~\ref{figure6} and \ref{figure7} give the typical results of the normalized localization length and the Hall conductivity for $m_z$=1.3, 1.48, 1.7 and 2.2.

For $m_z$=1.3, as shown in Figs.~\ref{figure6}(a) and \ref{figure7}(a), with increasing of the disorder strength $W_x$, the system also undergoes the phase transitions from 3D QAH to DM and NI phase, which is similar with those given in Fig.~\ref{figure3}(a) for the nonmagnetic disorder. For $m_z$=1.48, the system is in the 3D QAH phase and closes to the phase boundary of 3D QAH-DWSM [see Fig.~\ref{figure1}(a)]. According to the results in Figs.~\ref{figure6}(b) and \ref{figure7}(b), $d\Lambda$/$dL$ alters from a negative to positive value at $W_x$=1 and the Hall conductivity decreases from a quantized value $\sigma_{xy}$=$16e^2$/$8h$ to $\sigma_{xy}$=$14e^2$/$8h$, which demonstrates a transition from 3D QAH to DWSM phase. This attractive result has not been found in previous studies~\cite{CCZ2,Shapourian,CR3,SY2,WYJ}. We use SCBA to explain the novel transition. In contrast with the nonmagnetic disorder that diminishes $\tilde{m}_z$, instead, according to Eq.~(\ref{mz}), the $\sigma_x$ orbital disorder has a positive correction to $\tilde{m}_z$. With increasing of the disorder strength $W_x$, $\tilde{m}_z$ gradually increases and arrives at the value of 1.5, which is a phase boundary between 3D QAH and DWSM phase [see Fig.~\ref{figure1}(a)]). The bulk gap closes and a pair of Weyl nodes emerge at the zone boundary, inducing the 3D QAH-DWSM transition. Further increasing the disorder strength $W_x$, it can render the Weyl nodes to approach each other. The decrease of the distance between the Weyl nodes induce a plateau-to-plateau transition of the Hall conductivity from $\sigma_{xy}$=$14e^2$/$8h$ to $\sigma_{xy}$=$10e^2$/$8h$ [see Fig.~\ref{figure7}(b)]. Note that the Hall conductivity of the system is proportional to the distance of two Weyl nodes. Subsequently, the system experiences the phase transition from DM to NI phase with increasing of $W_x$.

Likewise, the system undergoes the phase transitions from DWSM to DM phase, then to NI phase as the disorder strength $W_x$ increases for the case $m_z$=1.7, as shown in Fig.~\ref{figure6}(c) and \ref{figure7}(c). This is similar with the result of the nonmagnetic disorder [see Fig.~\ref{figure4}(c)]. However, for $m_z$=2.2, with increasing of $W_x$, unlike the nonmagnetic disorder's case, the system directly enters the NI phase from DWSM phase without undergoing DM phase. This is another difference between the nonmagnetic and $\sigma_x$ orbital disorder [see Fig.~\ref{figure6}(d)].
\subsection{COMBINED EFFECT OF ORBITAL DISORDERS}\label{sec:IID}
\begin{figure}
  \centering
  \includegraphics[scale=0.6]{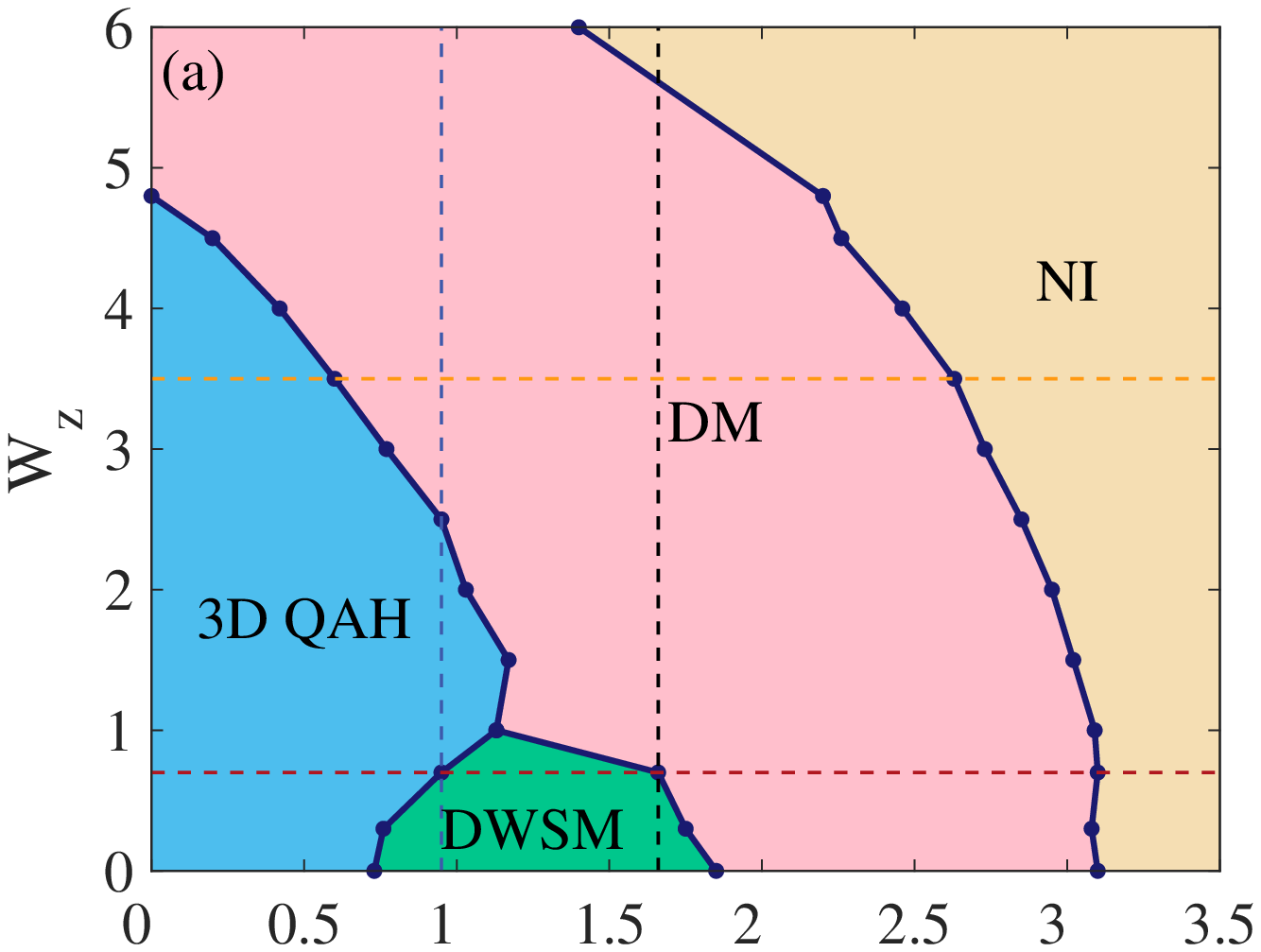}
  \includegraphics[scale=0.6]{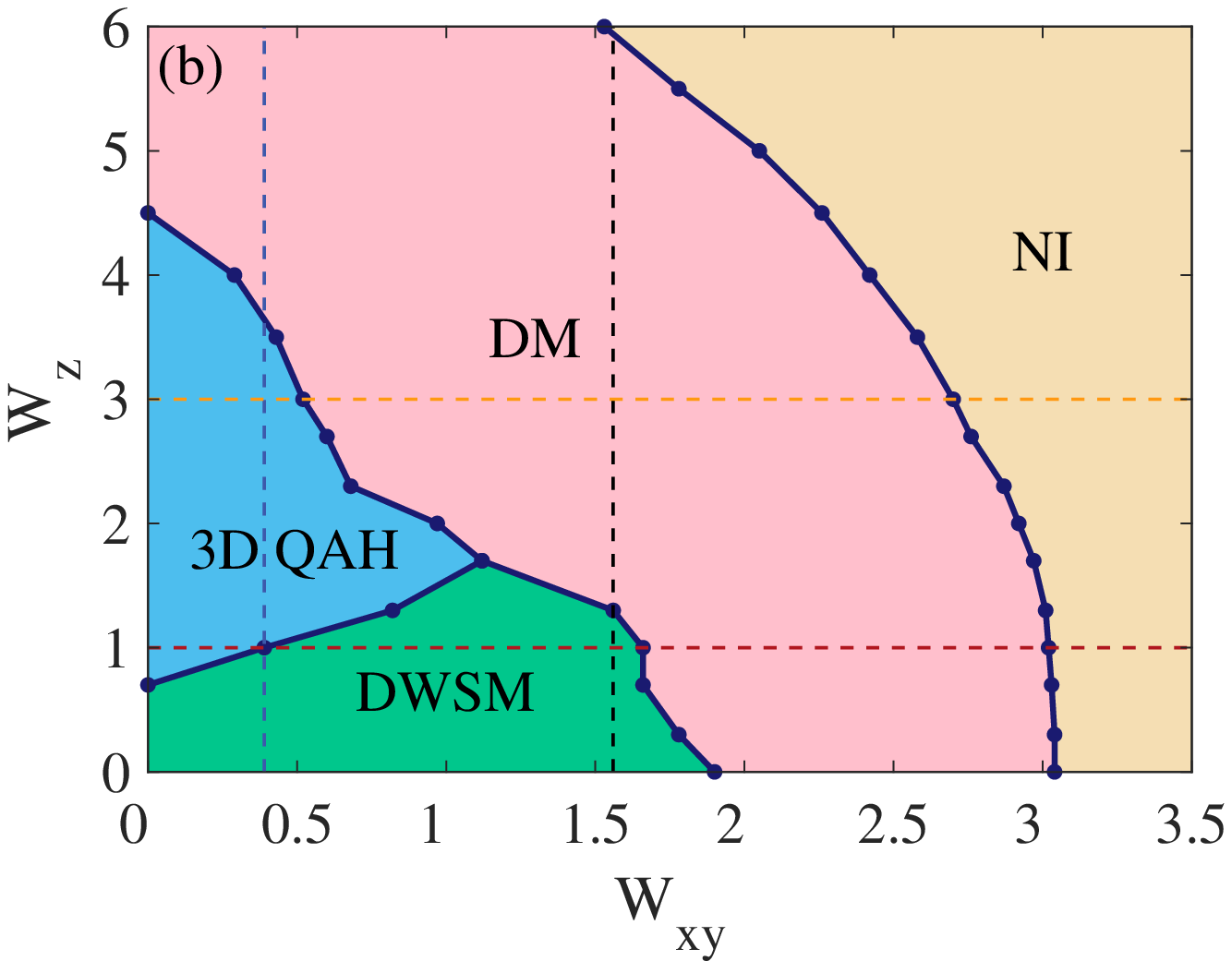}
  \caption{\label{figure8}Phase diagrams on the $W_z$-$W_{xy}$ plane for (a) $m_z$=1.48 and (b) $m_z$=1.52. The dark blue solid lines with dots are identified by the transfer matrix method. The accurate phases are determined by the Hall conductivity $\sigma_{xy}$. Other parameters are $t_z$=0.5 and $t_x$=$t_y$=1.}
\end{figure}

In this section, we discuss the combined effect of three kinds of the orbital disorders. Since $\sigma_x$ and $\sigma_y$ orbital disorder have identical correction to the mass term $m_z$, we take $W_x$=$W_y$=$W_{xy}$. We plot the phase diagrams on the $W_z$-$W_{xy}$ plane for $m_z$=1.48 [3D QAH phase] in Fig.~\ref{figure8}(a) and for $m_z$=1.52 [DWSM phase] in Fig.~\ref{figure8}(b).

For $m_z$=1.48, as shown in Fig.~\ref{figure8}(a), when $W_{xy}$$<$0.73, the system is first in 3D QAH phase, and then evolves into DM phase with increasing of $W_z$. When the disorder strength further increases, for example $W_{xy}$=0.95 [see the blue vertical dashed line], the increase of $W_z$ can induce the transition from DWSM to 3D QAH phase. We can interpret this phase transition according to the competition between $W_{xy}$ and $W_z$. This competition can be illustrated according to Eq.~(\ref{mz}). $W_{xy}$ has positive correction to $m_z$ while it's opposite for $W_z$. When $W_z$$<$0.7, the renormalized mass term $\tilde{m}_z$$>$1.5, so the system is still in DWSM phase. When $W_z$ exceeds 0.7, the system enters the 3D QAH phase because the effect caused by $W_z$ is greater than that of $W_{xy}$. The system goes into DM phase when $W_z$ further increases and reaches the value of 2.5. When the disorder strength increases to $W_{xy}$=1.66 [see the black vertical dashed line], the system undergoes the DWSM-DM-NI phase transitions. This transition happens because before being drove into the 3D QAH phase with increasing of $W_z$, the total disorder strength is strong enough that causes the system evolving into DM phase. Further increasing $W_z$, the system is localized and evolves into NI phase. For $W_z$=0.7 [see the red horizontal dashed line], when $W_{xy}$$>$0.95, the system enters DWSM phase from 3D QAH phase because the effect of $W_{xy}$ is greater than that of $W_z$. Further increasing $W_{xy}$ can render the system evolve into DM and NI phase in sequence. For the case $W_z$=3.5 [see the yellow horizontal dashed line], the system enters DM phase from 3D QAH phase when $W_{xy}$ surpasses 0.6. The reason is similar with that of DWSM-DM phase transition for the case $W_{xy}$=1.66.

We can use the same competition between $W_{xy}$ and $W_z$ to describe the phase transition for the case $m_z$=1.52, as shown in Fig.~\ref{figure8}(b). For $W_{xy}$=0.39 [see the blue vertical dashed line], the system experiences the DWSM-3D QAH-DM phase transitions with increasing of $W_z$. And for $W_{xy}$=1.56 [see the black vertical dashed line], the system undergoes the phase transition from DWSM to DM phase. For $W_z$=1.0 [see the red horizontal dashed line], the system enters DWSM phase  from 3D QAH phase when $W_{xy}$ exceeds 0.39. Then the system enters DM and NI phase with the increase of $W_{xy}$. For the case $W_z$=3.0 [see the yellow horizontal dashed line], a 3D QAH-DM phase transition happens with increasing of $W_{xy}$, and then the system evolves into NI phase when $W_{xy}$ exceeds the value of 2.76.

\section{SELF-CONSISTENT BORN APPROXIMATION}\label{sec:SCBA}
The self-consistent Born approximation is extensively used to analyze the weak disorder effect in various systems~\cite{Groth,SJT,Hsiang}. In the SCBA framework, the disorder can generate a self-energy correction to the Hamiltonian and therefore renormalize the model parameters. The self-energy $\Sigma$ can be calculated by the following integral equation:
\begin{eqnarray}\label{selfenergy}
\Sigma(E)=\frac{W^2}{12}\bigg(\frac{a}{2\pi}\bigg)^3\int_{BZ}d\textbf{\textit{k}}\{\sigma_i[E-h({\textbf{\textit{k}}})-\Sigma(E)]^{-1}\sigma_i\},
\end{eqnarray}
where the self-energy can be written as $\Sigma=\Sigma_0\sigma_0+\Sigma_x\sigma_x+\Sigma_y\sigma_y+\Sigma_z\sigma_z$. The renormalized mass term is $\tilde{m}_z$=$m_z$+$\Sigma_z$.
We consider the zeroth-order Born approximation, so we neglect $\Sigma$ in the right side of Eq.~(\ref{selfenergy}) and expand the Hamiltonian $h(\textbf{\textit{k}})$ at $(0,0,0)$ to $k^2$ order: $h(\textbf{\textit{k}})$$\approx$$\frac{1}{2}(k^2_y-k^2_x)\sigma_x+k_xk_y\sigma_y+[m_z+\frac{1}{2}(k^2_x+k^2_y+k^2_z)-3]\sigma_z$. Then we substitute $h(\textbf{\textit{k}})$ into Eq.~(\ref{selfenergy}) and get the renormalized mass term
\begin{eqnarray}\label{mz}
\tilde{m}_z=m_z+\frac{a^3W^2}{192\pi^2}\int_{-\pi}^{\pi}\ln\Bigg|\Bigg(\frac{\pi}{a}\Bigg)^4\frac{2}{(2m_z-6+k^2_z)^2}\Bigg|dk_z,
\end{eqnarray}
with $W^2$=$W_x^2$+$W_y^2$$-W_z^2$$-W_0^2$. The integration in the right side of Eq.~(\ref{mz}) is always positive, so the $\sigma_x$ orbital disorder $W_x$ and $\sigma_y$ orbital disorder $W_y$ enlarge the value of $\tilde{m}_z$, while the on-site nonmagnetic disorder $W_0$ and $\sigma_z$ orbital disorder $W_z$ diminish the $\tilde{m}_z$. The analysis from SCBA is consistent with the results calculated by the transfer matrix method.
\section{SUMMARY AND DISCUSSIONS}\label{sec:Discussion}
In summary, we study the effect of on-site nonmagnetic and orbital disorders on DWSM by calculating the localization length and the Hall conductivity. These disorders can give rise to rich phase transitions shown in the phase diagrams. Firstly, the tight-binding Hamiltonian of DWSM is introduced in momentum and real space. The phase diagram in the clean limit is presented. In the presence of disorder, the results indicate that the 3D QAH phase and DWSM phase are stable in the weak disorder. And the system undergoes a series of phase transitions with increasing of the disorder strength. For nonmagnetic disorder, phase transitions in DWSM are consistent with those reported in Weyl semimetal. However, for the $\sigma_x$ orbital disorder, it is the first time for us to find a new phase transition from 3D QAH to DWSM. Besides, there is a directly phase transition from DWSM to NI phase, which is quite different from the effect of the nonmagnetic disorder. For the latter, the system must enter the DM phase before being localized by the strong disorder.  The combined effect of orbital disorders are also investigated for disordered 3D QAH and DWSM phases. At last, an effective medium theory based on self-consistent Born approximation is introduced to explain the disorder-induced phase transitions.

The DWSM is realized experimentally in photonic crystal by using planar fabrication technology with the robustness of the surface state~\cite{CWJ}. Besides, the disorder-induced topological phase transition is observed experimentally from a trivial insulator to a topological Anderson insulator with robust chiral edge states~\cite{GGL}. The researchers fabricate a microwave-scale photonic crystal consisting of dielectric and gyromagnetic pillars. Disorder is introduced by randomly rotating the dielectric pillars in each unit cell, with the disorder strength parametrized by the maximum rotation angle. Therefore, we can utilize the same method to verify the disorder effect discussed in this article.

\begin{acknowledgments}
We would like to thank Professor You-Quan Li (Zhejiang University) and Professor Pei Wang (Zhejiang Normal University) for helpful discussions.
We thank the High-performance Computing Platform of Hangzhou Dianzi University for providing the computing resources and this work was supported by National Natural Science Foundation of China (Grant No. 11574067).
\end{acknowledgments}

\end{document}